 \documentclass{cjour}
 \usepackage[dvips]{graphicx}
 \newcommand{ \Vec   }[1]{ \mbox{\boldmath $ #1 $} }
 \newcommand{ \PD    }[2]{ \frac{ \partial #1 }{ \partial #2 } }

\begin{document}

%
 \journame{Journal of Computational Physics}
 \articlenumber{jcph.2002.7053}
 \yearofpublication{2002}
 \volume{179}
 \cccline{}
%
%
%
 \authorrunninghead{Shu-ichiro Inutsuka}
 \titlerunninghead{SPH with Riemann Solver}
%



\title{Reformulation of Smoothed Particle Hydrodynamics \\
       with Riemann Solver
       \thanks{ 
                to appear in 
                {\em Journal of Computational Physics}, 
                Vol. {\bf 179}, pp.238-267, 2002
               }
       }


\authors{Shu-ichiro Inutsuka}

\affil{Department of Physics, Kyoto University, 
       Kyoto, 606-8502, Japan 
       }

\email{inutsuka@tap.scphys.kyoto-u.ac.jp}

\abstract{
Smoothed Particle Hydrodynamics is reformulated in terms of 
the convolution of the original hydrodynamics equations, and 
the new evolution equations for the particles are derived.  
The same evolution equation of motion is also derived 
using a new action principle.  
The force acting on each particle is determined by solving the 
Riemann problem. 
The use of the Riemann Solver strengthens the method, 
making it accurate for the study of phenomena with strong shocks. 
The prescription for the variable smoothing length is shown. 
These techniques are implemented in strict conservation form. 
The results of a few test problems are also shown. 
}

\begin{article}

\section{Introduction}

Smoothed particle hydrodynamics (SPH, \cite{Lucy:SPH,GM:SPH}) 
is a fully Lagrangian particle method 
to describe the hydrodynamical phenomena.  
The Lagrangian particle methods are especially 
suited to hydrodynamical problems 
that have large empty regions and moving boundaries.    
Those problems naturally arise in engineering science 
as well as geophysics and astrophysics.  
A variety of astrophysical problems have been studied by SPH 
because of its simplicity in programming the two- and 
three-dimensional codes and its versatility of incorporating 
the various physical effects such as self-gravity, radiative cooling, 
and chemical reactions.   
A broad discussion of the method can be found 
in a review by Monaghan \cite{Monaghan:ARAA}.  
However, the inconsistency of the method is emphasized by Dilts
\cite{Dilts:1999,Dilts:2000} who modified the method by means of 
the moving-least-square basis functions to obtain accurate 
derivatives regardless of the positions of the SPH particles.      
Another concern of the method is its poor description 
of the strong shocks.  
In the two- or three-dimensional calculation of colliding gases, 
particles often penetrate into the opposite side.  
This unphysical effect can be partially eliminated by the so-called 
XSPH prescription \cite{Monaghan:1989} 
that does not introduce the (required) additional dissipation 
but results in the additional dispersion of the waves.  
Therefore it is very desirable to construct a method 
that is simple and able to describe the strong shock 
phenomena accurately.  

In this paper, a new method for handling shocks in 
particle hydrodynamics is constructed.  
The force acting on each particle is determined by solving 
the Riemann problem (RP). 
This use of the so-called Riemann Solver is 
introduced as an simple analogy of the grid-based method 
\cite{Godunov:1959}.  
The previous attempts to introduce Riemann Solver into the particle 
method failed to give the method an exact conservation form
\cite{SI:GPH,SI:IAU95}. 
This paper describes how to include the exact Riemann Solver into 
the strictly conservative particle method. 

The alternative approach to including small but sufficient 
dissipation into the numerical solution is to find a good 
limiter to switch a dissipative method and a less-dissipative 
method \cite{FulkQuinn:1995,MorrisMonaghan:1997}.  
In principle, however, the switch is always an option to any
numerical scheme including the present one, 
and its discussion is beyond the scope of this paper.

Section 2 provides a description of the method 
where we derive the evolution equations for the particles 
in terms of the convolution of the original hydrodynamic equations 
with the so-called kernel function. 
The same evolution equations are derived from 
an action principle which is different from the previous ones 
\cite{GM:1982, Nelson:1994}, 
which sheds light on the ``hidden'' approximation 
in the expression for the velocity field in SPH formalism.  
The detailed explanation for the implementation is 
described in Section 3.  
The usage of the Riemann Solver is analogous to 
the grid-based second-order Godunov scheme \cite{vL:1979}. 
A variable smoothing length is also considered. 
Numerical examples involving strong shocks are presented in Section 4. 
Section \ref{Sec:Summary} is for summary. 

\section{The Method}
We consider the following set of equations for 
non-radiating inviscid fluid: 
\begin{equation}
                  \frac{ d \rho }{ dt }
                             = - \rho \nabla \cdot \Vec{v} , 
                                                          \label{eq:EoC}
\end{equation}
\begin{equation}
                  \frac{ d \Vec{v} }{ dt } =
                             - \frac{1}{\rho} \nabla P , 
                                                          \label{eq:EoM}
\end{equation}
\begin{equation}
                  \frac{ d u }{ dt } =
                             - \frac{P}{\rho} \nabla \cdot \Vec{v} , 
                                                          \label{eq:EoE}
\end{equation}
\begin{equation}
                  P = ( \gamma - 1 ) \rho u,
                                                          \label{eq:EoS}
\end{equation}
where $u$ denotes the specific internal energy, $\gamma$ denotes 
the ratio of specific heats, and 
other symbols have their usual meanings.

In the standard SPH method, 
each particle has its own mass, and density at each particle's 
location is simply assigned by 
\begin{equation}
                 \rho_{i} =
                 \sum_{j} m_{j} W(\Vec{x}_{i} - \Vec{x}_{j} , h ), 
\end{equation}
where subscripts denote particle labels, 
$m_{j}$ is the mass of the $j$-th particle, 
and $W(\Vec{x},h)$ is a spherically symmetric kernel function 
which is normalized to be unity if integrated in space: 
\begin{equation}
                      W(\Vec{x},h) = W(-\Vec{x},h) , 
                                                  \label{eq:symm}
\end{equation}
\begin{equation}
                 \int W(\Vec{x},h) d\Vec{x} = 1 ,  
                                                  \label{eq:norm}
\end{equation}
and $h$ is the parameter of the kernel function. 
For later convenience we also define the effective width 
$h_{\rm eff}$ of the kernel function $W(\Vec{x})$ by 
\begin{equation}
                  h_{\rm eff}^2 \equiv 
                  2 \int x^2 W(\Vec{x},h) d\Vec{x} . 
                                                       \label{eq:heff}
\end{equation}
Although there are many possible forms for $W(\Vec{x},h)$, 
we use the following Gaussian kernel throughout in this paper: 
\begin{equation}
                 W(\Vec{x},h) =
                        \left[ \frac{1}{ h \sqrt{\pi} } \right]^{d}
                        e^{ - \Vec{x}^2 / h^2 } .
                                                  \label{eq:gaussian}
\end{equation}
In the above equation, $d$ is the number of dimensions, 
and $h$ is the so-called smoothing length. 
Note that $h_{\rm eff} = h$ for this Gaussian kernel. 
We assume $h$ is constant in space in Sections 
\ref{Sec:Pre}-\ref{Sec:Riemann}. 
The spatially variable smoothing length is considered
in Section \ref{Sec:vh} and the subsequent Sections.

The standard procedure to derive sets of evolution equations 
for SPH has been previously described \cite{Monaghan:ARAA}. 
In the following sections, we show how new evolution equations are 
derived from the original hydrodynamics equations.  
We then show that our new equation of motion can be derived from 
the action principle for an appropriately defined Lagrangian function. 

\subsection{Key Concept}
                                                       \label{Sec:Pre}
In general any method of computational fluid dynamics 
inevitably brings errors into the solutions.   
Therefore the essential feature of the method can be characterized 
by how to introduce the errors into the solutions.  
For example, in many kinds of grid-based methods,  
the truncation of the Taylor-series expansion is the origin 
of the errors in the numerical solutions.   
In the SPH formalism 
we consider the convolution of 
the physical function $f(x)$ with the kernel function,   
\begin{equation}
             \langle f \rangle (\Vec{x}) \equiv  
             \int f(\Vec{x}') W(\Vec{x}'-\Vec{x},h) d\Vec{x}' . 
                                                     \label{eq:conv}
\end{equation}
We use the symbol $\langle \rangle$ to denote the convolution. 
Taylor-series expansion of $f$ around $\Vec{x}=\Vec{x}_i$ 
in Eq.(\ref{eq:conv}) shows that the difference of 
$\langle f \rangle (\Vec{x}_i)$ and $f(\Vec{x}_i)$ is 
the second-order in $h_{\rm eff}$:    
\begin{equation}
             \langle f \rangle (\Vec{x}_i) = 
                f(\Vec{x}_i) 
             + \frac{h_{\rm eff}^2}{4} \nabla^2 f(\Vec{x}) 
             + O(h_{\rm eff}^4) .
                                                  \label{eq:ConvError}
\end{equation}
Thus if we use $\langle f \rangle$ 
as an approximate solution for $f$, 
the errors of the second-order in $h_{\rm eff}$ are introduced 
by the convolution with the symmetric function $W$. 

We also have 
\begin{eqnarray}
             \langle \PD{f}{x} \rangle (\Vec{x}) & = &
      \int \PD{f(\Vec{x}')}{x'}   W(\Vec{x}'-\Vec{x},h) d\Vec{x}'
                                                          \nonumber \\ 
    & = & - \int f(\Vec{x}') \PD{}{x'} 
                 W(\Vec{x}'-\Vec{x},h) d\Vec{x}'
                                                          \nonumber \\ 
    & = &   \PD{}{x} \int f(\Vec{x}')  
                 W(\Vec{x}'-\Vec{x},h) d\Vec{x}', 
                                                     \label{eq:gconvn}
\end{eqnarray}
where we used the integration by part and Eq.(\ref{eq:symm}). 
That is, the kernel convolution of $\nabla f$ is just 
$\nabla \langle f \rangle $. 

If we define the density distribution by the summation of 
the kernel functions at particle positions: 
\begin{equation}
                 \rho(\Vec{x}) \equiv 
                 \sum_{j} m_j W(\Vec{x} - \Vec{x}_{j} , h ), 
                                                    \label{eq:Dsph}
\end{equation}
then we have the following identities: 
\begin{equation}
               1 = \sum_{j} \frac{m_j}{\rho(\Vec{x})} 
                              W(\Vec{x} - \Vec{x}_{j} , h ) , 
                                                     \label{eq:Id1}
\end{equation}
\begin{equation}
               0 = \sum_{j} m_{j} 
                   \nabla 
                   \left[
                   \frac{ W(\Vec{x}-\Vec{x}_{j},h) }{ \rho(\Vec{x}) }
                   \right] .
                                                       \label{eq:Id2}
\end{equation}
These equations are the key equations in the present method. 
Using Eq.(\ref{eq:Id1}) 
the value of the kernel convolution can be cast into 
the following expression: 
\begin{eqnarray}
    f_i \equiv \langle f \rangle (\Vec{x}_i) 
    & = &  \int f(\Vec{x}') W(\Vec{x}'-\Vec{x}_i,h) d\Vec{x}'
                                                          \nonumber \\ 
    & = &  \int \sum_j m_j \frac{ f(\Vec{x}') }{\rho(\Vec{x}')} 
                W(\Vec{x}'-\Vec{x}_i,h) 
                W(\Vec{x}'-\Vec{x}_j,h) d\Vec{x}'
                                                          \nonumber \\ 
    & = &   \sum_j \Delta f_{i,j}, 
                                                     \label{eq:convn}
\end{eqnarray}
where we define 
\begin{equation}
     \Delta f_{i,j} \equiv 
           \int m_j \frac{ f(\Vec{x}') }{\rho(\Vec{x}')} 
                W(\Vec{x}'-\Vec{x}_i,h) 
                W(\Vec{x}'-\Vec{x}_j,h) d\Vec{x}' .
                                                     \label{eq:fij}
\end{equation}
Note that $m_i \Delta f_{i,j}$ is symmetric 
with respect to $i$ and $j$. 
In this way the value ($f_i$) of the physical variable $f$ at the 
$i$-th particle position is expressed as the summation of 
the contributions ($\Delta f_{i,j}$) from the surrounding particles.  
This expression shows the spirit of the present method, 
although the actual detailed evolution equations are presented 
in the following sections. 

In contrast the standard SPH adopts the following form 
for $\Delta f_{i,j}$: 
\begin{equation}
     \Delta f_{i,j} \approx
                m_j \frac{ f(\Vec{x}_j) }{\rho(\Vec{x}_j)} 
                W(\Vec{x}_i-\Vec{x}_j,h). 
                                                     \label{eq:fijSPH}
\end{equation}
This corresponds to the crude approximation, 
$W(\Vec{x}_j-\Vec{x}',h) \approx \delta(\Vec{x}'-\Vec{x}_j)$ 
in Eq.(\ref{eq:fij}), 
where $\delta(\Vec{x})$ is the Dirac's $\delta$ funcion.  
The standard SPH also implicitly assumes 
the following approximate equation: 
\begin{equation}
                 1 \approx \sum_{j} \frac{m_j}{\rho(\Vec{x}_j)} 
                                    W(\Vec{x} - \Vec{x}_{j} , h ), 
                                                     \label{eq:SPH1}
\end{equation}
instead of Eq.(\ref{eq:Id1}),    
although this is very poor approximation.  

In the MLSPH scheme by Dilts, an equation analogous to 
Eq.(\ref{eq:Id1}) is provided by 
the moving-least-square basis functions,    
although he needed somewhat heavy operations 
to construct his basis functions. 
Note that our method does not restrict the functional form of the 
kernel function to satisfy Eqs.(\ref{eq:Dsph}),(\ref{eq:Id1}).   
%
%
%
 \subsection{Equation of Motion}  
                                                 \label{sec:relation}

To derive the appropriate evolution equation for particles
we take the convolution of the equation of motion (EoM), 
Eq.(\ref{eq:EoM}), 
\begin{equation}
       \int \frac{ d \Vec{v}(\Vec{x}) }{ dt } 
                                W(\Vec{x}-\Vec{x}',h) d\Vec{x} =
     - \int \frac{1}{\rho(\Vec{x})} \nabla P(\Vec{x}) ~
                                W(\Vec{x}-\Vec{x}',h) d\Vec{x}.
                                               \label{eq:EocEoM}
\end{equation}
The right-hand side of this equation can be transformed into 
the following expression: 
\begin{eqnarray}
 &&
     - \int \left\{ 
                      \left[ \nabla  \frac{P(\Vec{x}) }{\rho(\Vec{x})}
                      \right]
                             W(\Vec{x}-\Vec{x}',h) 
                    +        \frac{P(\Vec{x}) }{\rho^2(\Vec{x})} 
                      \left[ \nabla \rho(\Vec{x}) 
                      \right]
                             W(\Vec{x}-\Vec{x}',h) 
            \right\}          d\Vec{x}  
                                                        \nonumber \\
 &&  = 
       \int \left\{ 
                             \frac{P(\Vec{x}) }{\rho(\Vec{x})} 
                      \nabla~W(\Vec{x}-\Vec{x}',h) 
                    -        \frac{P(\Vec{x}) }{\rho^2(\Vec{x})} 
                      \left[ \nabla~\sum_{j} m_j 
                                    W(\Vec{x}-\Vec{x}_{j},h)
                      \right]
                             W(\Vec{x}-\Vec{x}',h) 
            \right\}          d\Vec{x}  
                                                        \nonumber \\
 &&  = \sum_{j} m_{j} 
       \int \frac{P}{\rho^2}
              \left\{ 
                      \nabla W(\Vec{x} -\Vec{x}' ,h)
                             W(\Vec{x} -\Vec{x}_j,h)
                    -        W(\Vec{x} -\Vec{x}' ,h)
                      \nabla W(\Vec{x} -\Vec{x}_j,h)
              \right\} 
                                                         d\Vec{x}
                                                        \nonumber \\
                                                \label{eq:EoM3}
\end{eqnarray}
where we integrated by part and 
used the identity Eq.(\ref{eq:Id1}). 
 
Thus if we adopt the following equation for the evolution 
of the particle positions,  
\begin{equation}
       \ddot{\Vec{x}}_i
   \equiv   \int \frac{ d \Vec{v}(\Vec{x}) }{ dt } 
                                W(\Vec{x}-\Vec{x}_i,h) d\Vec{x} 
   = - \int \frac{1}{\rho(\Vec{x})} \nabla P(\Vec{x}) ~
                                W(\Vec{x}-\Vec{x}_i,h) d\Vec{x} ,   
                                                \label{eq:EoM4}
\end{equation}
we obtain the evolution equation 
that is consistent and spatially second-order accurate 
to the original hydrodynamical equation of motion: 
\begin{equation}
              m_i \ddot{\Vec{x}}_i  = 
              - m_i \sum_{j} m_j 
                         \int \frac{ P(\Vec{x}) }{ \rho^2(\Vec{x}) }
                         \left[  \PD{}{\Vec{x}_i} 
                               - \PD{}{\Vec{x}_j}  \right]
                         W(\Vec{x}-\Vec{x}_i, h)
                         W(\Vec{x}-\Vec{x}_j, h) d\Vec{x} , 
                                                        \label{eq:EoM2}
\end{equation}
where the overdot 
means a time-derivative.  
The antisymmetric appearance of $i$ and $j$ on the right-hand side 
guarantees the conservation of the linear and angular momentum of 
the particle system. 
From this equation we can deduce some properties 
of the EoM of the particle system.  
For example, if the pressure distribution is constant in space, 
the acceleration vanishes exactly in Eq. (\ref{eq:EoM2}), 
which is evident in Eq. (\ref{eq:EoM4}).  
The evolution equation of MLSPH also has 
a similar property although the standard SPH does not 
\cite{Dilts:1999}.

In this way the main approximation introduced in the present method 
is simply expressed in Eq. (\ref{eq:EoM4}), 
while the standard SPH formalism does not have such simple 
explanation.

 \subsection{Energy Equation}
                                                   \label{Sec:Energy}
We can follow the evolution of barotropic fluid with $P=P(\rho)$ 
only by the EoM derived in the previous section. 
However an additional evolution equation for energy 
is required to describe the evolution of non-barotropic fluid.  
The energy equation is also needed to handle shocks.  
The energy equation guarantees the conservation of the total energy 
of the system in the absence of other sources or losses of energy, 
such as radiative heating or cooling.  
On the other hand, 
the strict conservation property in the numerical calculation 
guarantees the accurate description of strong shocks.  
This is because the structure of shock wave is determined by 
the Rankine-Hugoniot relation which is the direct consequence 
of the physical conservation laws. 
Thus we must derive the energy equation that has 
strict conservation property and a convenient form 
to include additional physical dissipation.

As in Eq. (\ref{eq:EocEoM}), 
we multiply both sides of Eq. (\ref{eq:EoE}) by 
$W(\Vec{x}-\Vec{x}', h)$, 
and integrate in space (with respect to $\Vec{x}$), 
\begin{equation}
       \int \frac{ d u(\Vec{x}) }{ dt } 
                                W(\Vec{x}-\Vec{x}',h) d\Vec{x} =
     - \int \frac{P(\Vec{x})}{\rho(\Vec{x})} [\nabla \cdot \Vec{v}]~
                                W(\Vec{x}-\Vec{x}',h) d\Vec{x}. 
                                               \label{eq:EocEoE}
\end{equation}
The right-hand side of this equation can be transformed into 
the following expression: 
\begin{eqnarray}
 &   &  - \int \frac{P(\Vec{x})}{\rho(\Vec{x})} [\nabla \cdot \Vec{v}]~
                                W(\Vec{x}-\Vec{x}',h) d\Vec{x}   
                                                      \nonumber \\
 & = &  - \int \frac{1}{\rho(\Vec{x})} [\nabla \cdot P \Vec{v}]~
                                W(\Vec{x}-\Vec{x}',h) d\Vec{x}   
        + \int \frac{1}{\rho(\Vec{x})} [\Vec{v} \cdot \nabla P]~
                                W(\Vec{x}-\Vec{x}',h) d\Vec{x}  
                                                      \nonumber \\
\end{eqnarray}
At this point, we again make an approximation according to 
the following equation 
(see also Eqs.(\ref{eq:difvv}),(\ref{eq:evf})): 
                                       
\begin{eqnarray}
          \int \frac{1}{\rho(\Vec{x})} [\Vec{v} \cdot \nabla P]~
                                W(\Vec{x}-\Vec{x}_i,h) d\Vec{x}   
    & = & \int \frac{1}{\rho(\Vec{x})} [\dot{\Vec{x}}_i \cdot \nabla P]~
                                W(\Vec{x}-\Vec{x}_i,h) d\Vec{x}   
          + O(h^2).  
                                                      \nonumber \\
\end{eqnarray}
Then we have 
\begin{eqnarray}
            \dot{u_i}
 & \equiv  &
            \int \frac{du(\Vec{x})}{dt} W(\Vec{x}-\Vec{x}_i,h) d\Vec{x}
                                                          \nonumber \\
 & \approx & - \int \frac{1}{\rho(\Vec{x})} [\nabla \cdot P \Vec{v}]~
                                W(\Vec{x}-\Vec{x}_i,h) d\Vec{x}   
      +   \int \frac{1}{\rho(\Vec{x})} [\dot{\Vec{x}}_i \cdot \nabla P]~
                                W(\Vec{x}-\Vec{x}_i,h) d\Vec{x}  
                                                          \nonumber \\
 & = & \int P [ \Vec{v} - \dot{\Vec{x}}_i] 
          \cdot \nabla
                \left[ \frac{ W(\Vec{x}-\Vec{x}_i,h) }{\rho} \right]
                d\Vec{x}   
                                                          \nonumber \\
 & = & \sum_j m_j 
          \int \frac{ P }{\rho^2}
          [ \Vec{v} - \dot{\Vec{x}}_i ]  
          \cdot
              \left[ 
                      \nabla W(\Vec{x}  -\Vec{x}_i,h)
                             W(\Vec{x}  -\Vec{x}_j,h)
              \right. 
                                                          \nonumber \\
 &   & ~~~~~~~~~~~~~~~~~~~~~~~~~~~~~
              \left. 
                    -        W(\Vec{x}  -\Vec{x}_i,h)
                      \nabla W(\Vec{x}  -\Vec{x}_j,h)
              \right] 
              d\Vec{x}  , 
                                                          \nonumber \\
                                                        \label{eq:EoE4}
\end{eqnarray}
where we have again used the identities (\ref{eq:Id1}), 
(\ref{eq:Id2}). 
This form of the energy equation will be further modified 
to account for the physical dissipation in Section \ref{Sec:Riemann}, 
and its conservation property will be shown in Section \ref{Sec:Cons}.

\subsection{Action Principle}
                                                     \label{Sec:AP}
In this section we show another feature of the equation of motion 
in the present method.  
In the case without dissipation, 
the equation of motion for the particles 
in our method can be derived from an action principle. 
Therefore the time evolution of the particle system can be 
formulated in a canonical transformation, 
that might further enable the possible improvement 
of the time integration. 

For the moment, we consider barotropic fluid in which pressure is 
a function of density alone, $P(\rho)=K \rho^{\gamma}$. 
In this case, the equation of motion (Eq.[\ref{eq:EoM}]) 
for the fluid can be derived by minimizing the action $S$  
which is the time integral of the Lagrangian function $L$ 
expressed in Lagrangian coordinates:
\begin{equation}
                  S = \int L dt , 
\end{equation}
\begin{equation}
                  L = \int {\cal L} d\Vec{x}
                    = \int \left[ \frac{1}{2} \rho \Vec{v}^2 - \rho u 
                           \right] d\Vec{x} . 
                                                  \label{eq:L}
\end{equation}
If we use Eulerian coordinates, the appropriate Lagrangian density 
$\cal L$ must include constraint equations 
along with Lagrange multipliers \cite{Lin:1963,Serrin:1959},
which introduce additional complication. 
In contrast, the Lagrangian density shown above is simple and analogous 
to the classical particle mechanics form, 
owing to the use of Lagrangian coordinates. 
In addition we can derive the Hamiltonian and the evolution becomes 
the canonical transformation which is free from dissipation. 
Although this formalism is for the continuous media, 
we can use this formalism to derive the evolution equation for 
the system of particles. 

First we consider the density field expressed in the 
following equation: 
\begin{equation}
                 \rho(\Vec{x}) = 
                 \sum_{j} m_{j} W(\Vec{x} - \Vec{x}_{j} , h ).
                                                    \label{eq:Dsph1}
\end{equation}
This continuous distribution of density is defined only by 
(the finite number of) the positions of particles, 
$\Vec{x}_i~(i=1,2,3,...,N_{\rm p})$. 
Time-derivative of the above equation gives 
\begin{equation}
              \PD{ \rho(\Vec{x}) }{ t } 
            =   \sum_{j} m_{j} \dot{ \Vec{x}}_j \cdot \PD{}{\Vec{x}_j}
                                    W(\Vec{x} - \Vec{x}_{j} , h ) 
            = - \sum_{j} m_{j} \dot{ \Vec{x}}_j 
                             \nabla W(\Vec{x} - \Vec{x}_{j} , h )  
            = - \nabla \cdot \Vec{F}_{\rm m} , 
                                                   \label{eq:EoC2}
\end{equation}
where we defined 
\begin{equation}
            \Vec{F}_{\rm m}(\Vec{x}) \equiv 
                              \sum_{j} m_{j} \dot{ \Vec{x}}_j 
                              W(\Vec{x} - \Vec{x}_{j} , h ) . 
\end{equation}
Comparing Eq. (\ref{eq:EoC2}) with the ordinary continuity equation, 
we realize that $\Vec{F}_{\rm m}$ defines the mass flux, 
\begin{equation}
              \Vec{F}_{\rm m} (\Vec{x}) 
            = \rho(\Vec{x}) \Vec{v}(\Vec{x}) 
                                                    \label{eq:DoM}
\end{equation}
From this equation we can deduce the definition of velocity field 
given by the following equation: 
\begin{equation}
              \Vec{v}(\Vec{x})
            = \frac{ \Vec{F}_{\rm m}(\Vec{x}) }{\rho(\Vec{x})}
            = \frac{ \sum_{j} m_{j} \dot{ \Vec{x}}_j 
                              W(\Vec{x} - \Vec{x}_{j} , h ) }  
                   {\sum_{j} m_{j} W(\Vec{x} - \Vec{x}_{j} , h )} .
                                                     \label{eq:Dov}
\end{equation}
Note that at $\Vec{x}=\Vec{x}_i$ the above equation gives 
\begin{equation}
              \Vec{v}(\Vec{x}_i)
            = \dot{\Vec{x}}_i 
              + \frac{1}{\rho_i}
                \sum_{j} m_{j} [ \dot{\Vec{x}}_j - \dot{\Vec{x}}_i ]
                              W(\Vec{x}_i - \Vec{x}_{j} , h ) . 
                                                     \label{eq:XSPH}
\end{equation}
The difference of $\Vec{v}(\Vec{x}_i)$ and $\dot{\Vec{x}}_i$ is 
second order in $h$.  
In a one-dimensional case, for example, this can be shown 
by using an appropriate smooth function $f_x$ that satisfies 
$ f_x(x_i) = \dot{x}_i ~~(i=1,2,..., N_{\rm p})$ as 
%
%
\begin{eqnarray}
       &   & v_x (x_i) - \dot{x}_i 
                                                         \nonumber \\
       & = & 
                \frac{1}{\rho_i}
                \sum_{j} m_{j} 
          \left\{               \PD{   f_x}{x  }  [ x_j - x_i ]
                  + \frac{1}{2} \PD{^2 f_x}{x^2}  [ x_j - x_i ]^2 
                                             + O( [ x_j - x_i ]^3 )
          \right\}
                         W( x_i - x_j, h ) 
                                                         \nonumber \\
       & = & 
                \frac{1}{\rho_i}
                \frac{h^2}{2}
          \left\{               \PD{   f_x}{x  } \PD{\rho}{x}
                  + \frac{1}{2} \PD{^2 f_x}{x^2}     \rho
          \right\}
                + O( h^4 ) . 
                                                     \label{eq:difvv}
\end{eqnarray}

Eqs. (\ref{eq:Dsph1}) and (\ref{eq:Dov}) can be used for 
$\Vec{v}(\Vec{x})$ and $\rho(\Vec{x})$ in Eq. (\ref{eq:L}) 
to obtain the Lagrangian function 
that is defined only by the positions of particles.  
\begin{equation}
        L = 
            \int \left\{ 
            \frac{ \left[\sum_{j} m_{j} \dot{ \Vec{x}}_j 
                         W(\Vec{x} - \Vec{x}_{j}, h ) \right]^2 }  
                 {2 \sum_{j} m_{j} W(\Vec{x} - \Vec{x}_{j}, h )} 
            - \frac{K}{\gamma -1}
                 \left[
                    \sum_{j} m_{j} W(\Vec{x} - \Vec{x}_{j},h )
                 \right]^{\gamma}
            \right\} d\Vec{x} . 
                                                  \label{eq:L1}
\end{equation}
This is the exact Lagrangian for the ``fluid'' of which 
density and velocity are constrained by 
Eqs. (\ref{eq:Dsph1}) and (\ref{eq:Dov}). 

Instead of adopting this complicated Lagrangian function,  
we now make an approximation for the velocity field 
by Taylor-series expansion of the velocity field: 
\begin{eqnarray}
       \Vec{v}(\Vec{x}) & = & \Vec{v}(\Vec{x}_i) 
                  + [\Vec{x}-\Vec{x}_i] \cdot \nabla \Vec{v}(\Vec{x}_i)
                  + O(|\Vec{x}-\Vec{x}_i|^2)  
                                                   \\ \nonumber
       & = &    \dot{\Vec{x}}_i  
              + \frac{1}{\rho_i}
                \sum_{j} m_{j} [ \dot{\Vec{x}}_j - \dot{\Vec{x}}_i ]
                               W(\Vec{x}_i - \Vec{x}_{j} , h ) 
              + [\Vec{x}-\Vec{x}_i] \cdot \nabla \Vec{v}(\Vec{x}_i)
              + O(|\Vec{x}-\Vec{x}_i|^2) . 
                                                     \label{eq:evf}
\end{eqnarray}
From this we have 
\begin{equation}
       \int \Vec{v}(\Vec{x}) W(\Vec{x}-\Vec{x}_i,h) d\Vec{x}
                   = \dot{\Vec{x}}_i  
                     + O(h^2) . 
                                                     \label{eq:Aov}
\end{equation}
With this equation we can simplify the kinetic term of Lagrangian 
function as follows:
\begin{eqnarray}
                \int \frac{1}{2} \rho \Vec{v}^2 d\Vec{x}
  & = & \frac{1}{2} \int \Vec{F}_{\rm m} \cdot \Vec{v} d\Vec{x}
                                                   \nonumber \\
  & = & \frac{1}{2} \sum_{j} m_{j} \dot{\Vec{x}}_j \cdot
        \int \Vec{v}W (\Vec{x} - \Vec{x}_{j} , h ) d\Vec{x}
                                                   \nonumber \\
  & = & \frac{1}{2} \sum_{j} m_{j} \dot{\Vec{x}}_j^2 + O(h^2) .
\end{eqnarray}

Now we can define a simple Lagrangian function 
that is second-order accurate in $h$ as    
\begin{equation}
                L_{\rm new} \equiv \sum_i m_i \left[ 
                          \frac{1}{2} \dot{\Vec{x}}_i ^2 - 
                     \int u(\Vec{x}) W(\Vec{x}-\Vec{x}_i, h) d\Vec{x}
                                     \right] . 
                                                  \label{eq:Lgsph}
\end{equation}
We can derive the evolution equation for the particles by minimizing 
the action $ S = \int L_{\rm new}dt $ . 
In fact the Euler-Lagrange's equation, 
 \begin{equation}
                \frac{d}{dt} \PD{L_{\rm new}}{\dot{\Vec{x}}_i} 
                           - \PD{L_{\rm new}}{     \Vec{x}_i } = 0 ,
                                                        \label{eq:EL}
 \end{equation}
 gives 
\begin{equation}
                \ddot{\Vec{x}}_i  = 
              - \sum_{j} m_j 
                         \int \frac{ P(\Vec{x}) }{ \rho^2(\Vec{x}) }
                         \left[  \PD{}{\Vec{x}_i} 
                               - \PD{}{\Vec{x}_j}  \right]
                         W(\Vec{x}-\Vec{x}_i, h)
                         W(\Vec{x}-\Vec{x}_j, h) d\Vec{x} .
                                                       \label{eq:EoM2.2}
\end{equation}
The manipulation to derive Eq. (\ref{eq:EoM2.2}) from 
Eq. (\ref{eq:EL}) is explained in Appendix. 
This equation is the same as Eq. (\ref{eq:EoM2}).  

This is the exact evolution equation for the system of which Lagrangian 
function is defined by Eq. (\ref{eq:Lgsph}). 
The space-symmetry of $L_{\rm new}$ guarantees 
the conservation of linear momentum and angular momentum, 
which is also obvious in the anti-symmetric appearance of 
$i$ and $j$ in Eq. (\ref{eq:EoM2.2}). 

%
If we start with more simplified Lagrangian function, 
\begin{equation}
                L_{\rm SPH} = 
                \sum_i m_i \left[ \frac{1}{2} \dot{\Vec{x}}_i ^2 
                                  - u(\rho_i) 
                           \right] , 
                                                  \label{eq:Lsph}
\end{equation}
we would obtain the standard SPH equation \cite{GM:1982},
\begin{equation}
                \ddot{\Vec{x}}_i = 
              - \sum_{j} m_j \left[   
                                    \frac{ P_i }{ \rho_i^2 }
                                  + \frac{ P_j }{ \rho_j^2 } 
                             \right]
                \PD{}{\Vec{x}_i} W(\Vec{x}_i-\Vec{x}_j, h).
                                                        \label{eq:EoM1}
\end{equation}
%
This Lagrangian function $L_{\rm SPH}$ for the standard SPH can be 
obtained if we make a crude approximation, 
$W(\Vec{x}-\Vec{x}_i,h) \approx \delta(\Vec{x}-\Vec{x}_i)$ 
in Eq.(\ref{eq:Lgsph}).  
Thus the difference of the evolution equations, 
Eqs. (\ref{eq:EoM2.2}) and (\ref{eq:EoM1}) is related to the difference 
of the degrees of approximations 
to the original Lagrangian function of the fluid.  
In this context, the standard SPH is also assuming the approximation 
for the velocity field Eq.(\ref{eq:Aov}) as in the present method.  

Dilts \cite{Dilts:1999} reported that the SPH approximation is derived 
by means of the Galerkin approximation followed by 
the kernel approximation. 
In contrast to the Galerkin approximation, the above action principle 
has deep physical consequences such as the conservation of the linear 
and angular momentum of the particle system.   
In addition Hamiltonian structure of the method possibly gives further 
sophistication to the time-integration scheme   
(see, e.g, the symplectic integrator for the astronomical 
self-gravitating system \cite{WisdomHolman:1992}), 
although it is beyond the scope of this paper.

\section{Implementation}
                                                   \label{Sec:Imp}
In this section we describe the numerical implementation of 
the method in detail. 
In Section \ref{Sec:Conv} 
we describe how to evaluate the integrals 
in the basic evolution equations. 
Section \ref{Sec:Riemann} 
is for the introduction of Riemann Solver into the method.  
Section \ref{Sec:vh} 
shows the prescription for the variable smoothing length. 
In Section \ref{Sec:Cons} we show that the present method remains 
a fully conservative scheme even after the discretization in 
space and time. 

\subsection{Convolution}
                                                   \label{Sec:Conv}
In this subsection we describe how to evaluate the spatial integrals 
in Eq.(\ref{eq:EoM2}) and (\ref{eq:EoE4}). 
First we have to know the distribution of 
$\rho^{-2}(\Vec{x})$ to calculate 
the integrals.  
That is, we have to construct appropriate interpolation 
(or extrapolation) function for $\rho^{-2}(\Vec{x})$ around 
each pair ($i$ and $j$) of particles. 
 
For convenience, we define the $s$-axis which is along 
the vector $\Vec{x}_i-\Vec{x}_j$ and has the origin at 
$(\Vec{x}_i-\Vec{x}_j)/2$. 
We use $\Vec{s}_{\perp}$ to symbolically denote the components of 
the axes that are perpendicular to the $s$-axis, 
and define the $\Vec{s}$-coordinate system. 
The unit vector in the $s$-direction is 
 $
    \Vec{e}_{i,j} \equiv (\Vec{x}_i-\Vec{x}_j)/|\Vec{x}_i-\Vec{x}_j| .
 $
We set $\Delta s_{i,j} \equiv s_i-s_j=|\Vec{x}_i-\Vec{x}_j|$,  
where $s_i$ and $s_j$ denote 
the $s$-components of the positions, $\Vec{x}_i$ and $\Vec{x}_j$, 
respectively.  

We define the specific volume and its gradient as follows:
\begin{equation}
                 V(\Vec{x}) = \frac{1}{\rho(\Vec{x})} , 
\end{equation}
\begin{equation}
                 \nabla V(\Vec{x}) = -\frac{1}{\rho^2(\Vec{x})} 
                                      \nabla \rho(\Vec{x}) 
                 = - \frac{1}{\rho^2(\Vec{x})}
                     \sum_j m_j \nabla W(\Vec{x}-\Vec{x}_j).
\end{equation}
We will make approximate function for $V(\Vec{x})$ in the 
$\Vec{s}$-coordinates. 
In the following subsections,   
two sets of equations with the different order of 
accuracy are shown. 
%
\subsubsection{Linear Interpolation}
                                                     \label{Sec:Linear}
The most simple choice for the approximate function is the linear 
interpolation which is expressed as 
\begin{equation}
       V(s) =  \frac{1}{\rho(s)} = C_{i,j} s + D_{i,j} , 
                                                        \label{eq:VsL}
\end{equation}
where 
\begin{eqnarray}
 C_{i,j} & = & \frac{ V(\Vec{x}_i) - V(\Vec{x}_j) }
                    { \Delta s_{i,j}   }  , 
                                                            \nonumber \\
 D_{i,j} & = & \frac{ V(\Vec{x}_i) + V(\Vec{x}_j)}{2} , 
\end{eqnarray}
From this we obtained 
\begin{equation}
       V^2(s) =  \frac{1}{\rho^2(s)} = [ C_{i,j} s + D_{i,j} ]^2 , 
                                                        \label{eq:Vs2L}
\end{equation}
With this interpolation, we can calculate the integral 
as follows: 
\begin{equation}
     \int \rho^{-2}(\Vec{x})
             W(\Vec{x}-\Vec{x}_i, h)
             W(\Vec{x}-\Vec{x}_j, h) d\Vec{x} = 
          V_{i,j}^2 (h) W(\Vec{x}_i-\Vec{x}_j, \sqrt{2}h) ,
                                                    \label{eq:IntV2s} 
\end{equation}
where we defined 
\begin{equation} 
                     V_{i,j}^2 (h) \equiv
                     \frac{ 1}{4 } h^2 C_{i,j}^2 + D_{i,j}^2     . 
                                                    \label{eq:V2ijL} 
\end{equation}
The above calculation corresponds to an approximation based on 
the linear expansion of $1/\rho^2$ in the perpendicular direction 
to the vector $\Vec{x}_i-\Vec{x}_j$, that is,  
\begin{equation} 
       \rho^{-2}(\Vec{s}) =        V^2(s)
                            + \Vec{s}_{\perp} \cdot \nabla V^2(\Vec{s})
                            + O(|\Vec{s}_{\perp}|^2) . 
\end{equation}
Note that 
the integral of 
$\left[ \Vec{s}_{\perp} \cdot \nabla V^2(\Vec{s}) \right]$ 
in Eq. (\ref{eq:IntV2s}) vanishes identically 
owing to the symmetry of the kernel.  

Next, we consider the following integral for arbitrary function 
$f(x)$: 
\begin{equation}
       \int \frac{f(\Vec{x})}{ \rho^2(\Vec{x}) }
             W(\Vec{x}-\Vec{x}_i, h)
             W(\Vec{x}-\Vec{x}_j, h) d\Vec{x} .  \nonumber 
\end{equation}
When we have some interpolation for the function $f(\Vec{x})$ 
and $\rho^{-2}(\Vec{x})$ , 
we can define the weighted average $f^*_{i,j}$ by the 
following equation: 
\begin{equation}
                 \int \frac{f(\Vec{x})}{ \rho^2(\Vec{x}) }
                      W(\Vec{x}-\Vec{x}_i, h)
                      W(\Vec{x}-\Vec{x}_j, h) d\Vec{x} =
                 f^*_{i,j} 
                 \int \frac{1}{ \rho^2(\Vec{x}) }
                      W(\Vec{x}-\Vec{x}_i, h)
                      W(\Vec{x}-\Vec{x}_j, h) d\Vec{x}  . 
\end{equation}
A straight-forward calculation of the above equation with 
the linear approximation for the function 
$f(\Vec{x})$ ($\approx s [f_i - f_j]/\Delta s_{i,j} + [f_i+f_j]/2$) 
and the specific choice (Eq.[\ref{eq:Vs2L}]) of the interpolation of 
$\rho^{-2}(\Vec{x})$ gives 
\begin{equation}
                f^*_{i,j} = \frac{ f_i - f_j }{\Delta s_{i,j} } 
                             s^*_{i,j}
                          + \frac{ f_i + f_j }{2}  , 
                                                     \label{eq:fijL}
\end{equation}
where 
\begin{equation}
          s^*_{i,j} = \frac{ h^2 C_{i,j}D_{i,j}        }
                           { 2 V^2_{i,j} } . 
                                                     \label{eq:sijL}
\end{equation}
That is, $f^*_{i,j}$ is the value of the (linearly approximated) 
function $f$ at the position $s^*_{i,j}$. 

The evolution equations now become the following:
\begin{equation}
                \ddot{\Vec{x}}_i  = 
              - 2 \sum_{j} m_j 
                         P^*V^2_{i,j} (h)
                         \PD{}{\Vec{x}_i} 
                         W(\Vec{x}_i-\Vec{x}_j, \sqrt{2}h) ,
                                                     \label{eq:EoM2.5}
\end{equation}
\begin{eqnarray}
        \dot{u_i}
          & = & - 2 \sum_j m_j 
           ( [P\Vec{v}]^* - P^* \dot{\Vec{x}}_i) 
                         V^2_{i,j} (h)
                         \PD{}{\Vec{x}_i} 
                         W(\Vec{x}_i-\Vec{x}_j, \sqrt{2}h) ,  
                                                     \label{eq:EoE2.5}
\end{eqnarray}
where 
the formal meanings of the $P^*$ and $[P\Vec{v}]^*$ are the 
values at the position $s_{i,j}^*$ of 
the linearly interpolated functions.  
In section \ref{Sec:Riemann}, however, 
we will change the values of these quantities by considering 
the physical dissipation. 

\subsubsection{Cubic Spline Interpolation}
                                                     \label{Sec:Spline}
Another convenient method for the approximation is 
based on the cubic spline interpolation of $\rho^{-1}$ 
along the $s$-axis : 
\begin{equation}
       V(s) =  \frac{1}{\rho(s)} = 
                   A_{i,j} s^3 + B_{i,j} s^2 + C_{i,j} s + D_{i,j} , 
                                                        \label{eq:VsS}
\end{equation}
where 
\begin{eqnarray}
 A_{i,j} & = &          -2 \frac{( V _i - V _j )}{(\Delta s_{i,j})^3}
                         + \frac{( V'_i + V'_j )}{(\Delta s_{i,j})^2},
                                                            \nonumber \\
 B_{i,j} & = & \frac{1}{2} \frac{( V'_i - V'_j )}{ \Delta s_{i,j}   },
                                                            \nonumber \\
 C_{i,j} & = & \frac{3}{2} \frac{( V _i - V _j )}{ \Delta s_{i,j}   }  
             - \frac{1}{4}       ( V'_i + V'_j )                     ,
                                                            \nonumber \\
 D_{i,j} & = & \frac{1}{2}       ( V _i + V _j ) 
             - \frac{1}{8}       ( V'_i - V'_j )   \Delta s_{i,j}    ,
                                                            \nonumber \\
    V_i  & = & V(\Vec{x}_i)  ,
                                                            \nonumber \\
    V_j  & = & V(\Vec{x}_j)  , 
                                                            \nonumber \\
    V_i' & = & \Vec{e}_{i,j} \cdot \nabla V(\Vec{x}_i) ,
                                                            \nonumber \\
    V_j' & = & \Vec{e}_{i,j} \cdot \nabla V(\Vec{x}_j) .
\end{eqnarray}
Then, 
\begin{equation}
       V^2(s) = \frac{1}{\rho^{2}(s)} = 
                \left[
                 A_{i,j} s^3 + B_{i,j} s^2 + C_{i,j} s + D_{i,j}
                \right]^2 , 
                                                       \label{eq:V2s}
\end{equation}
In this case Eq. (\ref{eq:IntV2s}) becomes 
\begin{equation}
     \int \rho^{-2}(\Vec{x})
             W(\Vec{x}-\Vec{x}_i, h)
             W(\Vec{x}-\Vec{x}_j, h) d\Vec{x} =   
     V_{i,j}^2 (h) W(\Vec{x}_i-\Vec{x}_j, \sqrt{2}h) ,
                                                    \label{eq:IntV2sC} 
\end{equation}
where we defined 
\begin{equation} 
       V_{i,j}^2 (h) \equiv
         \frac{15}{64} h^6   A_{i,j}^2 
       + \frac{ 3}{16} h^4 (2A_{i,j}C_{i,j}+B_{i,j}^2) 
       + \frac{ 1}{4 } h^2 (2B_{i,j}D_{i,j}+C_{i,j}^2) 
       +                     D_{i,j}^2     . 
                                                      \label{eq:V2ij} 
\end{equation}
Eq. (\ref{eq:sijL}) becomes 
\begin{equation}
          s^*_{i,j} = \frac{  \frac{15}{32} h^6  A_{i,j}B_{i,j} 
                            + \frac{3}{8}   h^4 (A_{i,j}D_{i,j}+
                                                 B_{i,j}C_{i,j}) 
                            + \frac{1}{2}   h^2  C_{i,j}D_{i,j} }
                           { V^2_{i,j} } . 
                                                      \label{eq:sijS} 
\end{equation}
To avoid undershooting and overshooting of the interpolating function, 
we need to use linear interpolation (Eq.[\ref{eq:VsL}]) in the case 
$V_i' V_j' <0 $.

 \subsection{The Usage of the Riemann Solver}
                                                   \label{Sec:Riemann}
The description of shock waves requires (physical) dissipative 
process which is not considered in the fundamental equations 
(Eqs.[\ref{eq:EoC}]-[\ref{eq:EoS}]). 
The Godunov's scheme and 
its second-order sequel (MUSCL, \cite{vL:1979}) and 
third-order sequel (PPM, \cite{CW:PPM}) 
use the exact Riemann Solver to include the (possibly) minimum and 
sufficient amount of dissipation into the method. 
At present these methods remain state-of-the-art grid-based methods 
for computational fluid dynamics. 
In this paper we introduce the exact and spatially second-order 
Riemann Solver into the particle method.

The Godunov's scheme uses the result of 
Riemann problem (RP) at each cell interface in the calculation 
of numerical flux (see, e.g., \cite{vL:1979}).  
%
%
Likewise, we want to use the result of RP at the vicinity of 
the middle point of the $i$-th particle and the $j$-th particle. 
This is achieved by modifying the values of $P^*$ and $[P\Vec{v}]^*$ 
in Eqs. (\ref{eq:EoM2.5}),(\ref{eq:EoE2.5}).  
The finite difference expression for this is the following:
\begin{eqnarray}
                \frac{\Delta \dot{\Vec{x}}_i}{\Delta t} 
              & = &
              - \sum_{j} m_j
                         P^*_{i,j}
                         \int \frac{1}{ \rho^2(\Vec{x}) }
                             \left[  \PD{}{\Vec{x}_i} 
                                   - \PD{}{\Vec{x}_j}  \right]
                         W(\Vec{x}-\Vec{x}_i, h)
                         W(\Vec{x}-\Vec{x}_j, h) d\Vec{x}
                                                          \nonumber \\
              & = &
              - 2 \sum_{j} m_j 
                         P^*V^2_{i,j} (h)
                         \PD{}{\Vec{x}_i} 
                         W(\Vec{x}_i-\Vec{x}_j, \sqrt{2}h) ,
                                                        \label{eq:EoM5}
\end{eqnarray}
\begin{eqnarray}
                \frac{\Delta u_i}{\Delta t} 
              & = &
               - \sum_{j} m_j
                     P^*_{i,j} ( \Vec{v}^*_{i,j} - \dot{\Vec{x}}^*_i ) 
                          \int \frac{1}{ \rho^2(\Vec{x}) }
                              \left[  \PD{}{\Vec{x}_i} 
                                    - \PD{}{\Vec{x}_j}  \right]
                          W(\Vec{x}-\Vec{x}_i, h)
                          W(\Vec{x}-\Vec{x}_j, h) d\Vec{x} 
                                                          \nonumber \\
              & = &
          2 \sum_j m_j 
           ( P^*\Vec{v}^* - P^* \dot{\Vec{x}}_i^*) 
                         V^2_{i,j} (h)
                         \PD{}{\Vec{x}_i} 
                         W(\Vec{x}_i-\Vec{x}_j, \sqrt{2}h) .  
                                                    \label{eq:EoE5}
\end{eqnarray}
where $\Delta$ stands for the finite difference of each variable, 
and $\dot{\Vec{x}}^*_i$ is the time-centered velocity of 
the $i$-th particle, 
and $P^*_{i,j}$ and $\Vec{v}^*_{i,j}$ are the results of RP 
between the $i$-th particle and the $j$-th particle.  
%
How to define and calculate these variables are 
explained below. 

\begin{figure}[htp]
\begin{center}
\includegraphics[width=7.2cm]{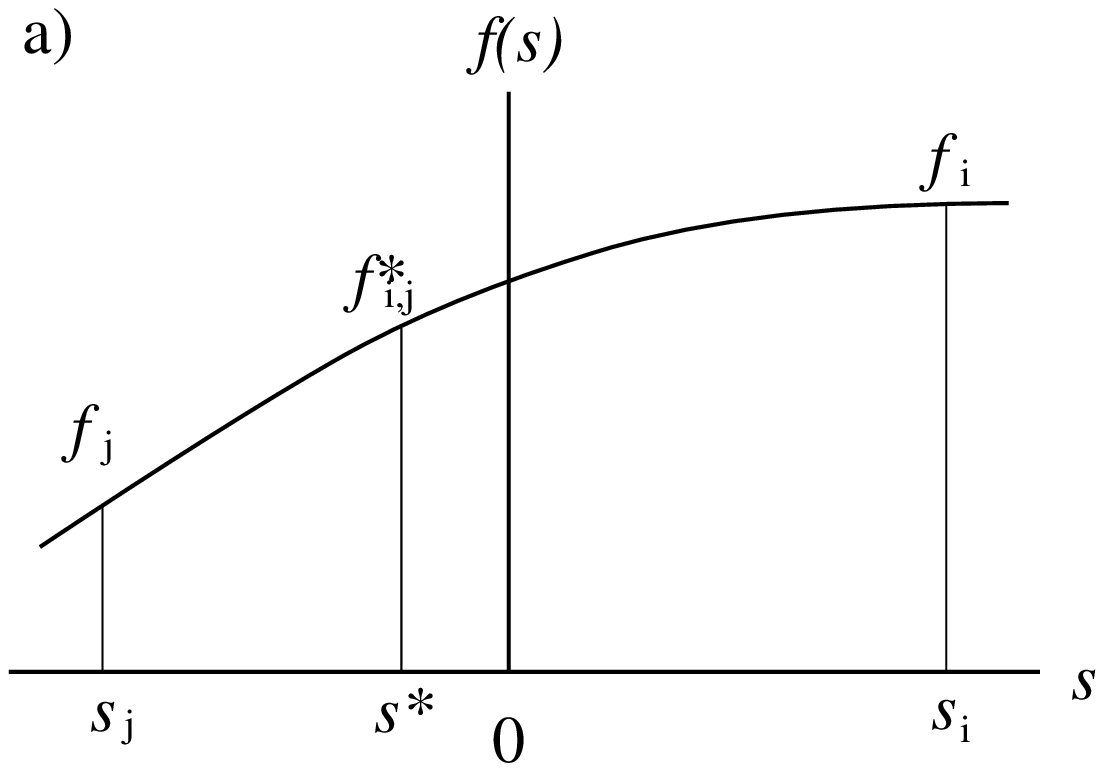}
\includegraphics[width=7.2cm]{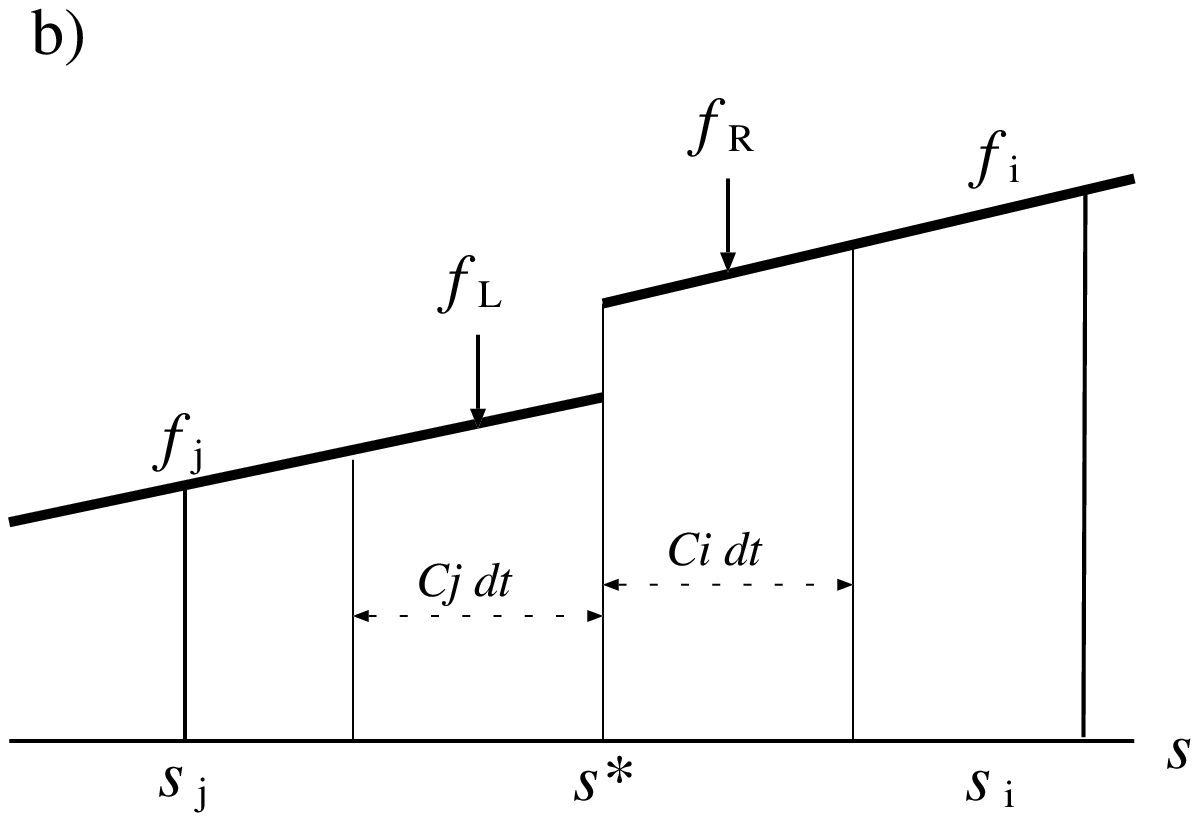}
\end{center}
   \caption{
a) A schematic picture of the distribution of the function $f(s)$ 
in the vicinity of the $i$-th particle and the $j$-th. 
The value of the weighted average $f^*_{i,j}$ 
are determined by Eqs. (\ref{eq:fijL}), 
and its location $s^*_{i,j}$ by 
Eq. (\ref{eq:sijL}) or (\ref{eq:sijS}). 
b) The setup of Riemann problem. 
We define the piecewise linear distribution of physical variable 
$f(s)$ in both the $i$-th region and the $j$-th region. 
The initial values of each side of the one-dimensional Riemann 
problem are the average values in each domain of dependence.
                                                       \label{Fig:RP}
}
\end{figure}

Figure (\ref{Fig:RP}) shows 
schematic picture of the distribution of the function $f(s)$ 
in the vicinity of the $i$-th particle and the $j$-th. 
According to the original (grid-based) Godunov's scheme, it is natural 
to define the interface of the two regions for the RP at $s^*_{i,j}$. 
To make the spatially second-order method like MUSCL, 
we define the piecewise linear distribution of physical variable 
$f(s)$ in both the $i$-th region and the $j$-th region. 
The gradient of $f(s)$ in each region can be simply assigned by 
the gradient at each particle's position. 
The initial values on each side of one-dimensional Riemann 
problem is the average values of each domain of dependence,  
which is expressed as follows: 
\begin{eqnarray}
       \rho_R & = &  \rho_i 
                  +  \left( \PD{\rho}{s} \right)_{i}
                     \left[ 
                      s^*_{i,j} + C_{{\rm s},i} \frac{\Delta t}{2} - s_i
                     \right],
                                                      \nonumber \\
       \  P_R & = &     P_i 
                  +  \left( \PD{   P}{s} \right)_{i}
                     \left[ 
                      s^*_{i,j} + C_{{\rm s},i} \frac{\Delta t}{2} - s_i
                     \right],
                                                      \nonumber \\
       \  v_R & = &     v_i 
                  +  \left( \PD{\  v}{s} \right)_{i}
                     \left[ 
                      s^*_{i,j} + C_{{\rm s},i} \frac{\Delta t}{2} - s_i
                     \right],
                                                      \nonumber \\
       \rho_L & = &  \rho_j 
                  +  \left( \PD{\rho}{s} \right)_{j}
                     \left[ 
                      s^*_{i,j} - C_{{\rm s},j} \frac{\Delta t}{2} - s_j
                     \right],
                                                      \nonumber \\
       \  P_L & = &     P_j 
                  +  \left( \PD{   P}{s} \right)_{j}
                     \left[ 
                      s^*_{i,j} - C_{{\rm s},j} \frac{\Delta t}{2} - s_j
                     \right],
                                                      \nonumber \\
       \  v_L & = &     v_j 
                  +  \left( \PD{   v}{s} \right)_{j}
                     \left[ 
                      s^*_{i,j} - C_{{\rm s},j} \frac{\Delta t}{2} - s_j
                     \right],
\end{eqnarray}
where 
\begin{eqnarray}
                  v_i & = & \Vec{v}_i \cdot \Vec{e}_{i,j} ,  
                                                      \nonumber \\
                  v_j & = & \Vec{v}_j \cdot \Vec{e}_{i,j} . 
\end{eqnarray}
and $C_{{\rm s},i}$ denotes the sound speed at $\Vec{x}_i$.
Thus we can solve the RP at the interface $s^*_{i,j}$. 
The detailed explanation for the solution of RP can be found in 
\cite{vL:1979,CW:PPM}, and we will not repeat it here. 

After solving the RP at $s^*_{i,j}$, we have the resulting pressure 
$P^*_{i,j}$ and the projected velocity $v^*_{i,j}$. 
Three-dimensional (or two-dimensional) velocity 
$\Vec{v}^*_{i,j}$ is calculated as follows: 
\begin{equation}
                \Vec{v}^*_{i,j} = v^*_{i,j} \Vec{e}_{i,j} + 
                          [ \Vec{v}_{i,j} - v_{i,j} \Vec{e}_{i,j} ],  
                                                       \label{eq:vaij}
\end{equation}
where 
\begin{equation}
                \Vec{v}_{i,j} =  \Vec{v}_i [1/2+\epsilon]
                               + \Vec{v}_j [1/2-\epsilon] ,  
                                                       \label{eq:v3ij}
\end{equation}
\begin{equation}
                v_{i,j} =  v_i [1/2+\epsilon] + v_j[1/2-\epsilon] ,  
                                                       \label{eq:vsij}
\end{equation}
\begin{equation}
                 \epsilon = \frac{ s^*_{i,j} }{ \Delta s_{i,j} } . 
                                                     \label{eq:epsilon}
\end{equation}
These $P^*_{i,j}$ and $\Vec{v}^*_{i,j}$ are used in the  
calculation of the right-hand side of Eqs. 
(\ref{eq:EoM5}) and (\ref{eq:EoE5}). 
Note that Eqs.(\ref{eq:v3ij}), (\ref{eq:vsij}), (\ref{eq:epsilon}) 
are not needed in the actual calculation, because 
the bracketed second term on the right-hand side of 
Eq. (\ref{eq:vaij}) is perpendicular to $\Vec{e}_{i,j}$ 
and vanishes identically in Eq.(\ref{eq:EoE5}). 

In the higher-order grid-based methods (e.g., MUSCL and PPM) 
we need to impose monotonicity constraint on the gradients of 
physical variable to obtain the stable description of the 
discontinuity.  
This is also true in the present method. 
Our experience shows that 
the monotonicity constraint only on the velocity field 
suffices to make the stable method. 
Thus in the actual calculation of our method, 
we mimic the monotonicity constraints by setting 
\begin{equation}
                  \left( \PD{v}{s} \right)_{i} =
                  \left( \PD{v}{s} \right)_{j} = 0
                  ~~~~
                  {\rm if}
                  ~~~
                  \left( \PD{v}{s} \right)_{i} \times
                  \left( \PD{v}{s} \right)_{j} < 0 . 
\end{equation}

The numerical scheme should be first-order in space at the surface 
of the shock wave. 
The shock surface tends to include a few particles in the 
actual calculation. 
Therefore, 
if the velocity difference of a certain particle pair corresponds to 
the sound speed divided by a small number, 
the pair is considered to be in the shock surface, 
and we need to use the first-order Riemann Solver for 
this pair.  
We implement this condition by setting 
\begin{equation}
                  \left( \PD{f}{s} \right)_i = 
                  \left( \PD{f}{s} \right)_j = 0
                  ~~~
                  {\rm if}
                  ~~
                  C_{\rm shock}~\Vec{e}_{i,j} \cdot 
                                ( \Vec{v}_j - \Vec{v}_i ) 
                  > \min(C_{{\rm s},i},C_{{\rm s},j}),  
\end{equation}
where $f=\rho,~P,~v$, 
and $C_{\rm shock}$ is a numerical constant corresponding to 
the number of particles at the shock surface.   
We adopt $C_{\rm shock}=3$ throughout in this paper.


\subsection{Variable Smoothing Length}
                                                   \label{Sec:vh}
In the previous subsections, we assume that 
the smoothing length, $h$, is constant in space.  
In actual calculation we need to change $h$ to enlarge the dynamic 
range of the spatial resolution. 
Therefore we have to extend the present method for the variable 
smoothing length. 

The formal derivation of the evolution equation with 
variable smoothing length is the same as Section \ref{sec:relation}, 
except that we should start with the definition of 
density as 
\begin{equation}
               \rho(\Vec{x}) = 
               \sum_{j} m_{j} W(\Vec{x} - \Vec{x}_{j} , h[\Vec{x}] ).
                                                      \label{eq:Dsph2}
\end{equation}
This definition of density corresponds to the so-called ``gather'' 
formulation of SPH \cite{HK:1989}. 
The resulting evolution equations are the following:
\begin{eqnarray}
                \ddot{\Vec{x}}_i & = & 
                \sum_{j} m_j
                         P^*_{i,j} 
                         \int \frac{1}{ \rho^2(\Vec{x}) }
                         \left\{ 
                                 \nabla W(\Vec{x}-\Vec{x}_i,h[\Vec{x}] )
                                        W(\Vec{x}-\Vec{x}_j,h[\Vec{x}] )
                         \right.
                                             \nonumber \\
            & & ~~~~~~~~~~~~~~~~~~~~~~~~~~~  
                         \left.
                               -        W(\Vec{x}-\Vec{x}_i,h[\Vec{x}] )
                                 \nabla W(\Vec{x}-\Vec{x}_j,h[\Vec{x}] )
                         \right\} 
                         d\Vec{x} , 
                                                        \label{eq:EoM6}
\end{eqnarray}
%
\begin{eqnarray}
                \dot{u}_i &  = &
                \sum_{j} m_j
                   P^*_{i,j} ( \Vec{v}^*_{i,j} - \Vec{v}_i )  
                   \int \frac{1}{ \rho^2(\Vec{x}) }         
                        \left\{ 
                                  \nabla W(\Vec{x}-\Vec{x}_i,h[\Vec{x}])
                                         W(\Vec{x}-\Vec{x}_j,h[\Vec{x}])
                         \right.
                                             \nonumber \\
            & & ~~~~~~~~~~~~~~~~~~~~~~~~~~~~~  
                         \left.
                                -        W(\Vec{x}-\Vec{x}_i,h[\Vec{x}])
                                  \nabla W(\Vec{x}-\Vec{x}_j,h[\Vec{x}])
                        \right\} 
                   d\Vec{x}. 
                                                    \label{eq:EoE6}
\end{eqnarray}
These equations are essentially the same as Eqs. 
(\ref{eq:EoM5}) and (\ref{eq:EoE5}),   
although the analytic integration for these equation is not possible 
even with the polynomial approximation for $\rho^{-2}(\Vec{x})$. 
Therefore we prefer to use simple approximation for the integral 
as in the following form: 
\begin{eqnarray}
             \frac{\Delta \dot{\Vec{x}}_i}{\Delta t} 
       & = &
              -   \sum_{j} m_j P^*
           \left[ ~
             V^2_{i,j} (h_i) 
             \PD{}{\Vec{x}_i} W(\Vec{x}_i-\Vec{x}_j, \sqrt{2}h_i)
           \right.
                                                        \nonumber \\
       &   & ~~~~~~~~~~~~~~~~~~~
           \left.
           + V^2_{i,j} (h_j) 
             \PD{}{\Vec{x}_i} W(\Vec{x}_i-\Vec{x}_j, \sqrt{2}h_j) ~
           \right] ,   
                                                     \label{eq:EoM6.5}
\end{eqnarray}
\begin{eqnarray}
                \frac{\Delta u_i}{\Delta t} 
       & = & - \sum_j m_j [ P^*\Vec{v}^* - P^* \dot{\Vec{x}}_i^* ] 
           \left[ ~
             V^2_{i,j} (h_i) 
             \PD{}{\Vec{x}_i} W(\Vec{x}_i-\Vec{x}_j, \sqrt{2}h_i)
           \right.
                                                        \nonumber \\
       &   & ~~~~~~~~~~~~~~~~~~~~~~~~~~~~~~~  
           \left.
           + V^2_{i,j} (h_j) 
             \PD{}{\Vec{x}_i} W(\Vec{x}_i-\Vec{x}_j, \sqrt{2}h_j) ~
           \right] ,   
                                                     \label{eq:EoE6.5}
\end{eqnarray}
where, in spirit, 
we used $h_i$ for the half of the integration space which 
include $\Vec{x}_i$, and used $h_j$ for the other half. 

The above approximations 
assume that the $h(\Vec{x})$ should not vary much 
within the neighborhood of each particle. 
One possible way to determine the smoothing length with this 
constraint is the following: 
\begin{equation}
                 h_i = \eta \left[ \frac{m_i}{\rho^*_i} \right]^{1/d}, 
\end{equation}
where 
\begin{equation}
                 \rho^*_i  = 
                 \sum_{j} m_{j} 
                  W(\Vec{x}_{i} - \Vec{x}_{j} , h^*_i  ) , 
                  ~~~ h^*_i = h_i C_{\rm smooth}.    
                                                     \label{eq:Dsmooth}
\end{equation}
$\rho^*$ is more smooth than $\rho$ itself if $C_{\rm smooth} > 1$ .    
Numerical experiments shows that 
$\eta \simeq 1$ with $C_{\rm smooth} \simeq 2$ 
works fine in Section \ref{sec:example}.  
The effective number of neighbors around each particle 
depends on the ratio of the smoothing length and the mean separation 
of particles at that particle.  
For example, we can usually ignore the contribution from 
the $j$-th particle to the $i$-th particle 
if their distance $|\Vec{x}_i-\Vec{x}_j|$ is 
larger than $3 h_i$, because $\exp(-3^2) \approx 1.234 \times 10^{-4}$. 
Thus, in one-dimensional calculations, the number of neighbors for 
calculating Eq.(\ref{eq:Dsmooth}) is $6 \eta C_{\rm smooth}$ 
excluding the $i$-th particle itself.  
The number of neighbors becomes about $28 \eta^2 C_{\rm smooth}^2$ 
in two-dimensional calculations,  
and about $113 \eta^3 C_{\rm smooth}^3$ in three dimensional 
calculations.

\subsection{Conservation Property}
                                                   \label{Sec:Cons}
The final discretized form for the equation of motion 
for the $i$-th particle becomes the following: 
\begin{eqnarray}
                \Delta \dot{\Vec{x}}_i
       & = &
              - \Delta t 
                \sum_{j} m_j P^*_{i,j}
                \left[
                       V^2_{i,j}(h_i) \PD{}{\Vec{x}_i} 
                       W(\Vec{x}_i-\Vec{x}_j, \sqrt{2}h_i)
                \right.
                                                        \nonumber \\
       &   & ~~~~~~~~~~~~~~~~~~~~~~~~~~~~~
                \left.
                     + V^2_{i,j}(h_j) \PD{}{\Vec{x}_i} 
                       W(\Vec{x}_i-\Vec{x}_j, \sqrt{2}h_j).
                \right]
                                                         \label{eq:EoM7}
\end{eqnarray}
This calculation guarantees the conservation of the total momentum of 
the system,   
\begin{equation}
                \sum_{i} m_i \Delta \dot{\Vec{x}}_i = 0 ,  
\end{equation}
because the terms in the square bracket in Eq. (\ref{eq:EoM7}) 
are anti-symmetric in $i$ and $j$. 

The final form for the equation of energy for the $i$-th particle 
becomes the following: 
\begin{eqnarray}
                 \Delta u_i
       & = &
               - \Delta t 
                 \sum_{j} m_j
                   P^*_{i,j} ( \Vec{v}^*_{i,j} - \dot{\Vec{x}}^*_i ) 
            \left[
                   V^2_{i,j}(h_i)
                   \PD{}{\Vec{x}_i} W(\Vec{x}_i-\Vec{x}_j, \sqrt{2}h_i)
            \right.
                                                        \nonumber \\
       &   & ~~~~~~~~~~~~~~~~~~~~~~~~~~~~~~~~~~~~
            \left.
                 + V^2_{i,j}(h_j)
                   \PD{}{\Vec{x}_i} W(\Vec{x}_i-\Vec{x}_j, \sqrt{2}h_j)
            \right]
                                                         \label{eq:EoE7}
\end{eqnarray}
where we defined the time-centered velocity of the $i$-th particle, 
\begin{equation}
                \dot{\Vec{x}}^*_i = \dot{\Vec{x}}_i 
                                  + \frac{1}{2} \Delta \dot{\Vec{x}}_i .
                                                         \label{eq:tcv}
\end{equation}
This expression guarantees the conservation of the total energy 
of the system, because 
\begin{eqnarray}
     \lefteqn{ \Delta  \sum_{i} m_i 
                \left[ \frac{1}{2} \Vec{\dot{x}}_i^2 + u_i \right] 
             } 
                                                     \nonumber \\ 
      & = &   \sum_{i} m_i 
              \left\{ \frac{1}{2} 
                       [ \Vec{\dot{x}}_i + \Delta \Vec{\dot{x}}_i ]^2
                      + [ u_i + \Delta u_i ]
                      - \frac{1}{2} \Vec{\dot{x}}_i^2 
                      - u_i 
              \right\}                                
                                                     \nonumber \\ 
      & = &   \sum_{i} m_i 
              \left\{ 
                     \Delta \Vec{\dot{x}}_i 
                     \left[ \Vec{\dot{x}}_i + 
                            \frac{1}{2} \Delta \Vec{\dot{x}}_i 
                     \right]
                   + \Delta u_i
              \right\}                                \nonumber \\ 
      & = &    
            - \sum_{i} \sum_{j} m_i m_j
                   P^*_{i,j} \Vec{v}^*_{i,j}
            \left[
                   V^2_{i,j}(h_i)
                   \PD{}{\Vec{x}_i} W(\Vec{x}_i-\Vec{x}_j, \sqrt{2}h_i)
            \right.
                                                        \nonumber \\
      &   &  ~~~~~~~~~~~~~~~~~~~~~~~~~~~~~~~~
            \left.
                 + V^2_{i,j}(h_j)
                   \PD{}{\Vec{x}_i} W(\Vec{x}_i-\Vec{x}_j, \sqrt{2}h_j)
            \right]    
                                                       \nonumber \\
      & = &    0 .                                  
                                                      \label{eq:CoE}
\end{eqnarray}
Thus the present scheme conserves energy exactly. 
This is in contrast to the ordinary energy equation of the standard SPH 
which is only accurate in the first-order in time  
(see, e.g., \cite{Benz:1989, Monaghan:ARAA}). 
We also note that the conservation of energy does not require 
constant smoothing length, which is obvious in 
Eq. (\ref{eq:CoE}).   
This is because 
the expression of the total energy in the numerical calculation  
        $\sum_{i} m_i ( \frac{1}{2} \Vec{x}_i^2 + u_i )$ 
does not explicitly depends on the choice of smoothing length. 
In other words, a sudden change of smoothing length at any timestep 
does not change the numerical value of the total energy. 


\subsection{Overall Procedure}
                                                   \label{Sec:Proc}

In this section 
we summarize the actual procedures in the sequence of executions. 
The main loop of the time integration corresponds to Steps 1-4.

\noindent {\bf Step 0. Problem Setup} \\
\hspace*{0.5cm} 
We first setup the problem in the computer program.  
We appropriately place the particles to represent the density 
distribution that corresponds to the initial condition of the 
problem to solve.  
This may require some relaxation technique to find 
the appropriate positions of the particles \cite{Monaghan:ARAA}.  
We also determine $\rho_i$, $\nabla \rho_i$, and $h_i$ to start 
the main loop of the time integration. 
Various constants and the initial variables are calculated 
in this step. 
For example we determine the initial timestep $\Delta t$.

\noindent {\bf Step 1. Gradient Calculation} \\
\hspace*{0.5cm} 
We calculate the gradient of physical variables $P, \Vec{v}$ 
for the use in the Riemann Solver. 

\noindent {\bf Step 2. Source Term Calculation} \\
\hspace*{0.5cm} 
We calculate the RHS of Eqs.(\ref{eq:EoM7}),(\ref{eq:EoE7}) 
either by Eq.(\ref{eq:V2ijL}) or Eq.(\ref{eq:V2ij}). 
The subroutine for the Riemann Solver is called once for every pair of 
particles to calculate $P^*$ and $v^*$ 
in Eqs.(\ref{eq:EoM7}),(\ref{eq:EoE7}).  
%
%

\noindent {\bf Step 3. Time Evolution} \\
\hspace*{0.5cm} 
We update $\Vec{x}_i$, $\Vec{\dot{x}}_i$, and $u_i$ according to  
Eqs.(\ref{eq:tcv}),(\ref{eq:EoM7}),(\ref{eq:EoE7}) ,   
\begin{eqnarray}
                     \Vec{x}_i (t+\Delta t)  
          & = &      \Vec{x}_i (t)           +       \dot{\Vec{x}}_i^* 
                                                     \Delta t , 
                                                          \nonumber \\ 
                \dot{\Vec{x}}_i(t+\Delta t)  
          & = & \dot{\Vec{x}}_i(t)          + \Delta \dot{\Vec{x}}_i , 
                                                          \nonumber \\ 
                          u_i  (t+\Delta t)  
          & = &           u_i  (t)          + \Delta           u_i . 
                                                      \label{eq:Update}
\end{eqnarray}

\noindent {\bf Step 4. Density Updation} \\
\hspace*{0.5cm} 
According to the updated positions of particles, 
we update density distribution.  
The smoothing length of each particle is also updated.  
The timestep $\Delta t$ for the next integration is also determined. 
We turn to Step 1 for the next time integration. 

\section{Numerical Examples}
                                                  \label{sec:example}
The present method was tested on a variety of 1D, 2D, and 3D problems, 
a few of which are described below. 
Other sets of test calculations will be described elsewhere. 

In determining the amount of the timestep $\Delta t$, 
we have to consider the Courant condition that is in spirit similar to 
the Courant condition for the grid-based Lagrangian methods,  
\begin{equation}
                \Delta t = 
                 C_{\rm CFL}
                \min_i \left\{ \left. 
                              \left[ \frac{m_i}{\rho_i}
                              \right]^{1/d} 
                               \right/
                               C_{{\rm s},i}
                       \right\}. 
\end{equation}
The numerical experiments show that we can safely use 
$C_{\rm CFL} \approx 0.5$ in most of hydrodynamical problems. 
Note that we don't need to consider 
the (effective) diffusion timescale that is related to 
the artificial viscosity adopted by the standard SPH and other 
particle methods.  
Thus $\Delta t$ in the present method can be much larger 
than that in the other SPH methods. 

The following examples are calculated with the Fortran program 
in which the single precision real number is used. 
To accelerate the computation we use the data structure based on 
link lists \cite{Monaghan:1985}.

Five cycles of iterations in the Riemann Solver is sufficient 
for the following test problems. 



\subsection{Shock Tube}
\begin{figure}[htp] 
\includegraphics[width=14.4cm]{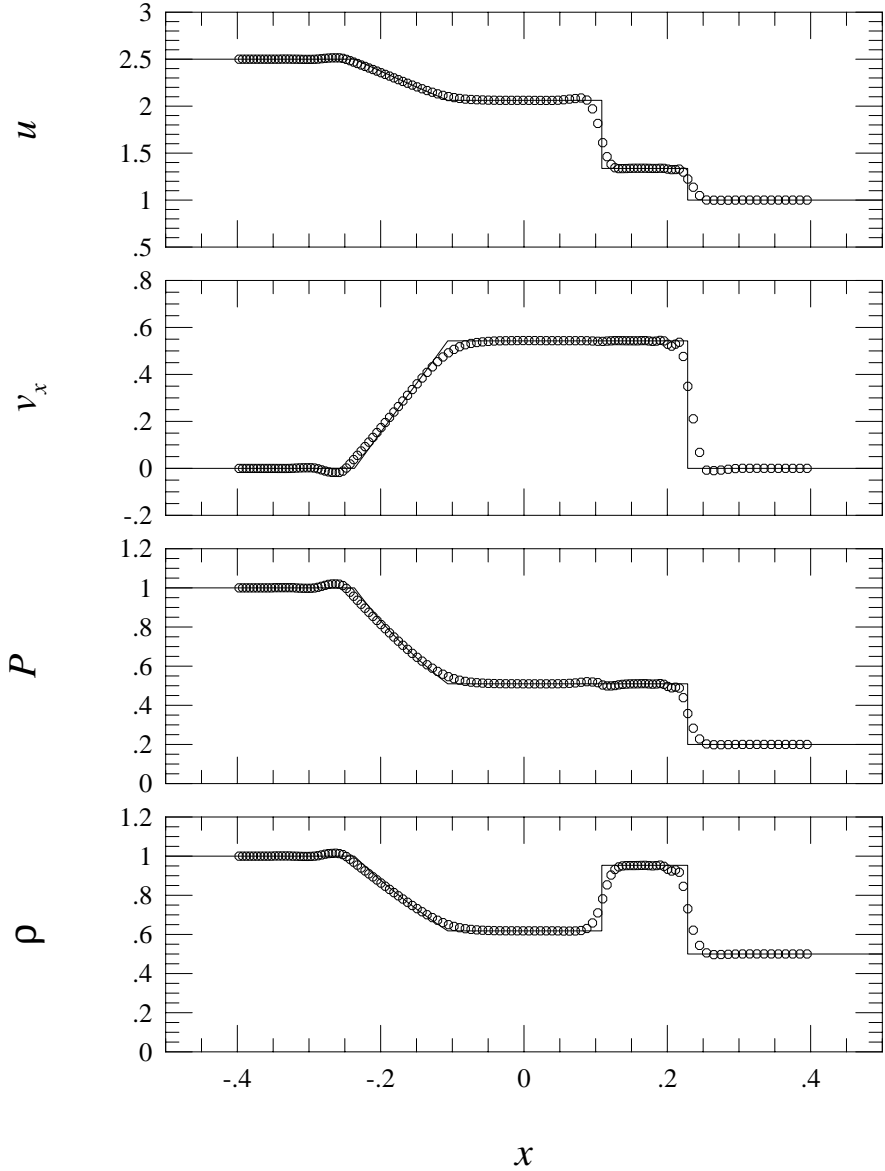}
   \caption
{
Results of the shock tube problem in which the Mach number is 1.526.
The right-hand side of the initial discontinuity 
includes 40 equal-mass particles.
The open circles plot the snapshot at $t=0.2$ 
by the present method with 
the cubic spline approximation for the convolution Eq.(\ref{eq:V2ij}) 
and the variable smoothing length ($\eta=1$, $C_{\rm smooth}=2.0$).
The solid lines correspond to the analytic solution.
                                                  \label{Fig:WSGSPH}
}
\end{figure}

\begin{figure}[htp] 
\includegraphics[width=14.4cm]{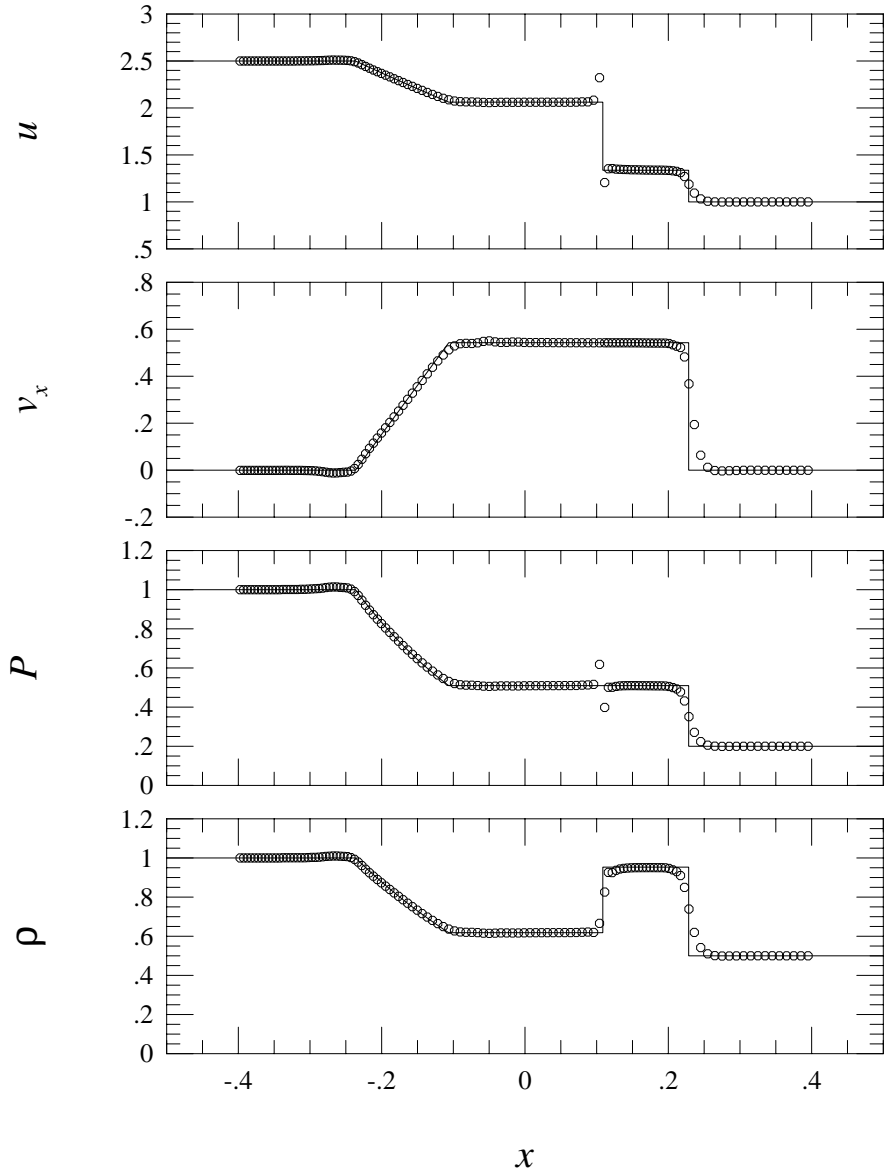}
   \caption
{
Results of the shock tube problem in which the Mach number is 1.526.
The open circles plot the snapshot at $t=0.2$ 
by the standard SPH. 
The solid lines correspond to the analytic solution.
                                                  \label{Fig:WSSPH}
}
\end{figure}

In order to compare the capabilities of the present method 
and standard SPH 
we test our method against the shock tube problems, 
of which the exact solutions are available. 
The density variation in the first problem 
is not so large that we can use both 
the variable smoothing length and 
the constant smoothing length. 
The initial parameters of the problem are the following:
\begin{eqnarray}
       \rho_{\rm L}  = 1 & , ~~~& \rho_{\rm R} = 0.5  , \nonumber \\
       P_{\rm L}     = 1 & , ~~~& P_{\rm R}    = 0.2    , \nonumber \\
       v_{x,{\rm L}} = 0 & , ~~~& v_{x,{\rm R}} = 0      ,
\end{eqnarray}
where the subscript ``L'' denotes the variable on the left-hand side
of the initial discontinuity, and R denotes the right-hand side.
The ratio of specific heats is $\gamma = 7/5$.  
The Mach number of the resultant shock wave is $1.526$.
The value of the post-shock pressure is $P^*=0.5099$. 
Figure \ref{Fig:WSGSPH} plots the snapshot at $t=0.2$ 
by the present method with 
the cubic spline approximation for the convolution Eq.(\ref{eq:V2ij}) 
and the variable smoothing length ($\eta=1$, $C_{\rm smooth}=2.0$).
The method with the linear approximation for the convolution 
Eq.(\ref{eq:V2ijL}) gives very similar result. 
Figure \ref{Fig:WSSPH} plots the corresponding result of 
the standard SPH 
where we used the ``standard'' artificial viscosity 
($\alpha=1,\beta=2$) described in 
the review paper by Monaghan \cite{Monaghan:ARAA}.  
In both cases, the number of equal-mass particles
is 80 (40) on the left(right)-hand side of initial discontinuity 
(i.e., $\Delta x_L=0.005$, $\Delta x_R=0.01$ ). 
The solid lines correspond to the analytic solution.
In this problem, these two methods give similar results 
except for the contact discontinuity
where the standard SPH produced a ``wiggle'' in pressure and 
specific internal energy.  
This is due to the inconsistency of EoM of the standard SPH 
(\ref{eq:EoM1}). 
The present method produced the smooth distribution of internal energy 
at the contact discontinuity, and hence provide the almost constant 
pressure distribution at the density discontinuity. 

The total energy of the system ($-0.4<x<0.4$) is defined as 
\begin{equation}
       \int_{-0.4}^{0.4} \frac{1}{2} \rho u~dx  = 1.2 . 
\end{equation}
The initial numerical value of the total energy of the particle system 
was the following: 
\begin{equation}
                \sum_{i} m_i u_i = 1.20000017
\end{equation}
where the last few digits have no significant meaning in this 
Fortran single precision calculation. 
The final ($t=0.2$) value of the total energy of the particle system 
was found to be 
\begin{equation}
                \sum_{i} m_i ( \frac{1}{2} \dot{\Vec{x}}_i^2 
                                       + u_i ) = 1.20003033
\end{equation}
The relative error remains sufficiently small ($\Delta E/E < 10^{-4}$)
when we change $\Delta t$ in the other runs.  
Thus the error of the total energy is only due to 
the round-off error of the single precision calculation, 
and not due to the truncation errors in the numerical modeling 
of the evolution.   
This result guarantees the strict conservation property of the 
present scheme (see Section \ref{Sec:Cons}).

The present method spent 0.17 seconds for the total 52 timesteps 
($\Delta t \approx 0.004$) to compute with 
the Hewlett-Packard workstation C240 (PA-RISC 8200/236MHz), 
which corresponds to $3.27\times 10^{-3}$ seconds per step and 
$2.72\times 10^{-5}$ seconds per particle. 
In contrast the standard SPH method used 956 timesteps 
($\Delta t \approx 0.0002$)
for the stable evolution in this problem, 
because the artificial viscosity limits $\Delta t$.  
Indeed numerical experiments shows larger $\Delta t$ causes unphysical 
oscillations in the solution. 
As a result the standard SPH method spent 1.35 seconds 
to finish the calculation.  
It corresponds to $1.4\times 10^{-3}$ seconds per step and 
$1.2\times 10^{-5}$ seconds per particle. 
This means that the present method is about two times slower than 
the standard SPH for the required operations per particle 
but the present method tends to spend much smaller (total) 
computation time than the standard SPH does. 

\begin{figure}[htp] 
\includegraphics[width=14.4cm]{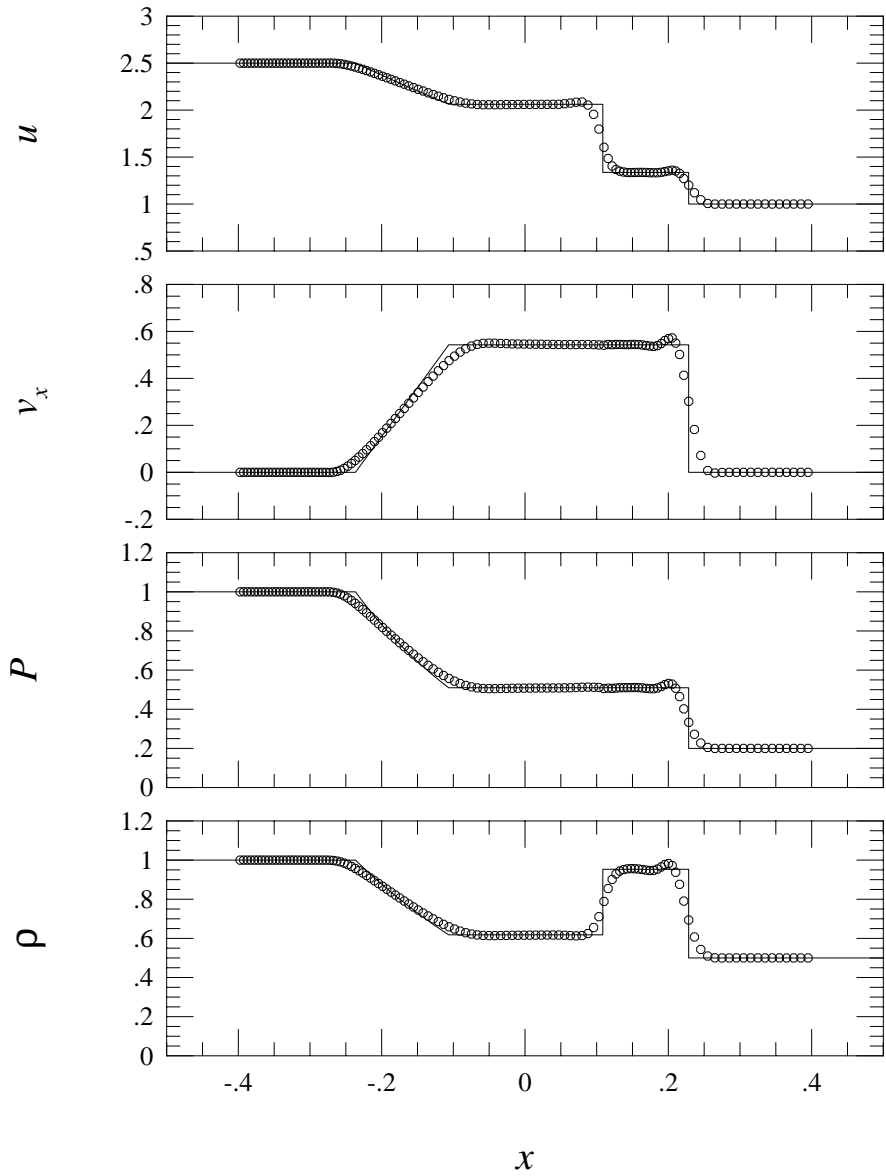}
   \caption
{
Same as Fig. 2 except that 
we used the constant smoothing length $h=0.01$. 
                                                 \label{Fig:WSGSPHch}
}
\end{figure}

Figure \ref{Fig:WSGSPHch} plot the result of the present method where  
we used the constant smoothing length $h=0.01$. 
The rarefaction wave on the left hand side is described 
less-accurately than in Fig. \ref{Fig:WSGSPH} as expected,   
because the smoothing length in high density region in 
Fig. \ref{Fig:WSGSPH} is smaller than the constant smoothing length 
in Fig. \ref{Fig:WSGSPHch}.  
The method with the constant smoothing length has no advantage 
over the method with the variable smoothing length 
even in this kind of shock tube problem where 
the density variation is not so large.  

\begin{figure}[htp] 
\includegraphics[width=14.4cm]{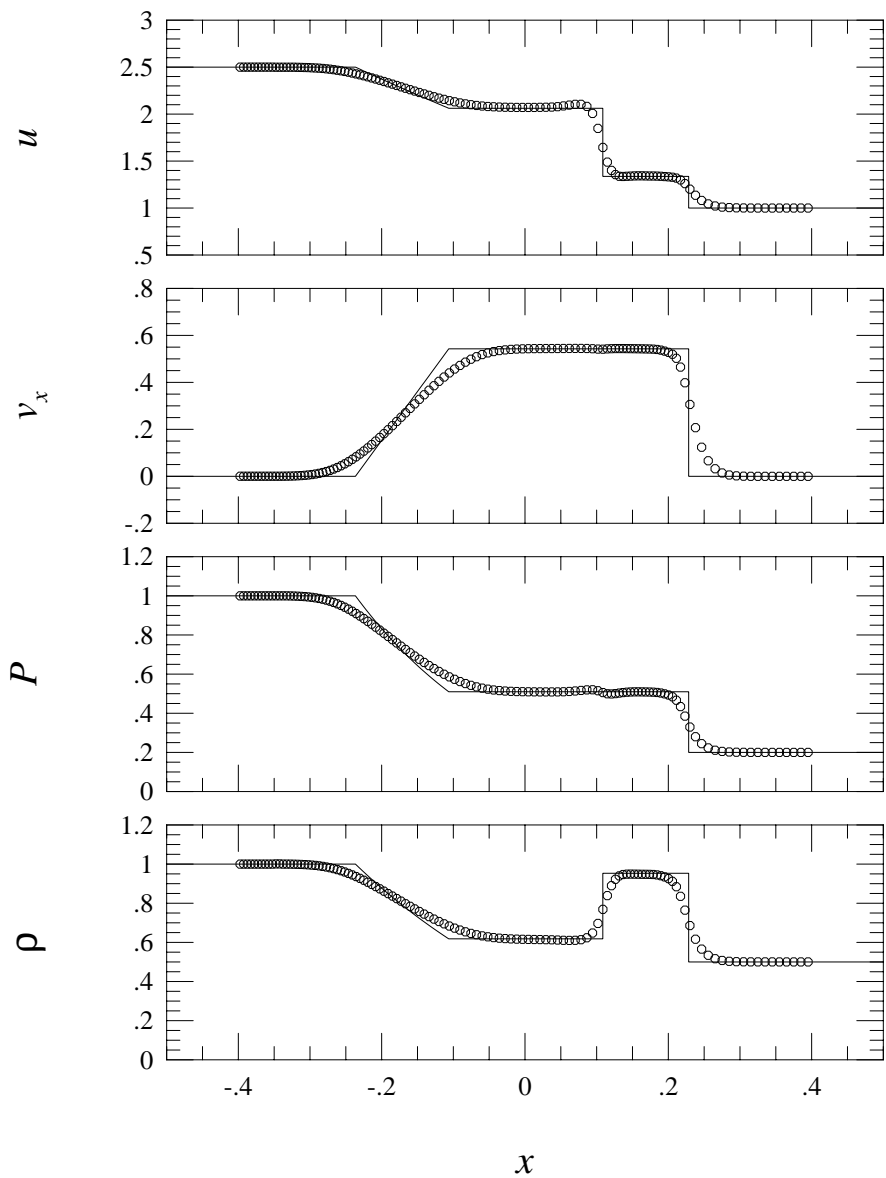}
   \caption
{
Same as Fig. 2 except that 
we used the first-order Riemann Solver.  
                                                 \label{Fig:WSGSPH1st}
}
\end{figure}

Figure \ref{Fig:WSGSPH1st} plot the result of the present method 
with the first-order Riemann Solver where we set 
$\partial \rho/\partial s=\partial P/\partial s
 =\partial v/\partial s=0$ 
in the Riemann Solver. 
The sharp profiles at the shock front and the head and tail of 
the rarefaction wave are smeared out as in the grid-based 
first-order Godunov method. 
In this calculation we don't need to calculate the gradients of 
$P$ and $v_x$.   
However the difference of the computation times with the first-
and second-order method is negligible. 
There is no reason to use this first-order method.

\begin{figure}[htp] 
\includegraphics[width=14.4cm]{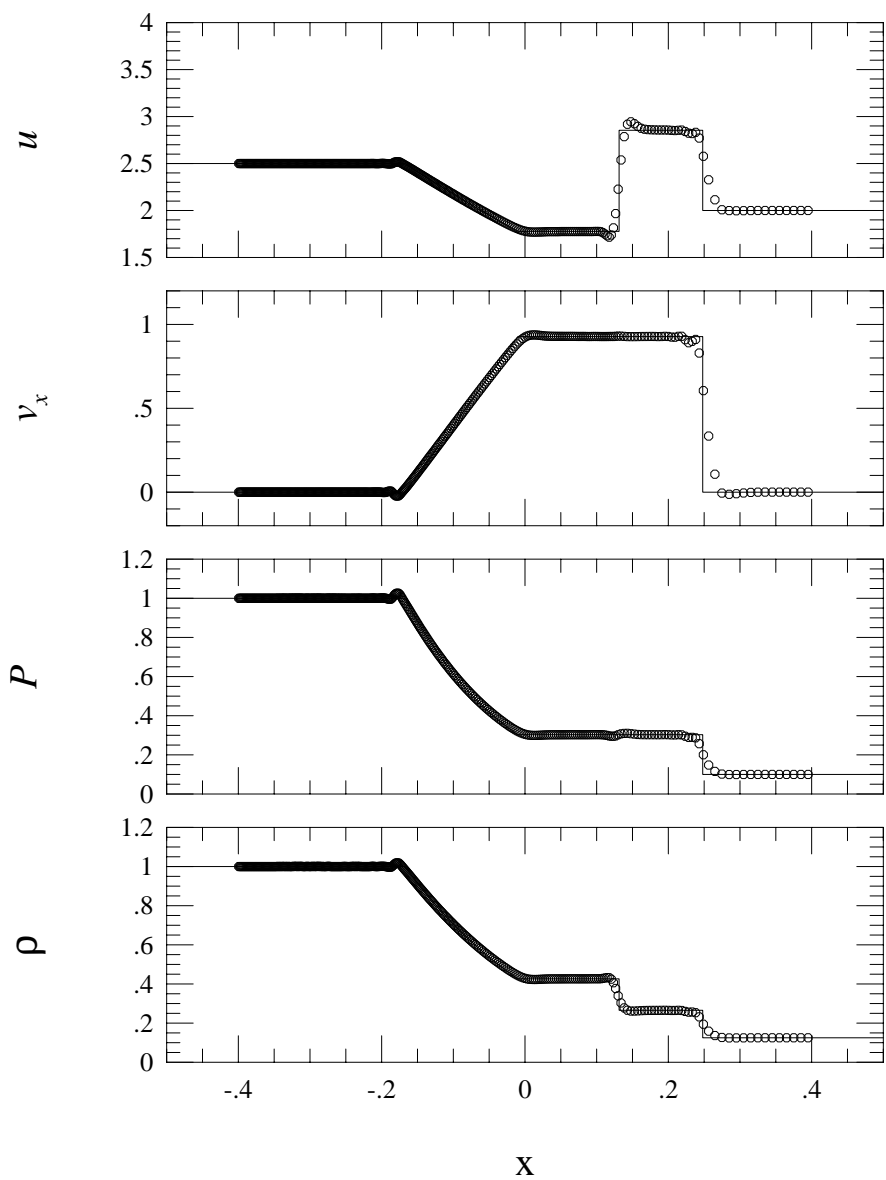}
   \caption
{
Results of the Sod's shock tube problem   
by the present method with 
the cubic spline approximation for the convolution Eq.(\ref{eq:V2ij}) 
and the variable smoothing length ($\eta=1$, $C_{\rm smooth}=2.0$).
The solid lines correspond to the analytic solution.
                                                       \label{Fig:Sod1}
}
\end{figure}

In Fig. \ref{Fig:Sod1},
the results of the standard Sod's shock tube \cite{Sod:1978} 
are plotted.
The initial parameters of the problem are the following:
\begin{eqnarray}
       \rho_{\rm L}  = 1 & , & \rho_{\rm R} = 0.125  , \nonumber \\
       P_{\rm L}     = 1 & , & P_{\rm R}    = 0.1    , \nonumber \\
       v_{x,{\rm L}} = 0 & , & v_{x,{\rm R}} = 0     .
\end{eqnarray}
The ratio of specific heats is $\gamma = 7/5$.
The solid lines correspond to the analytic solution.
The number of the equal-mass particles
in the $x$-direction is 40 on the right-hand side of the initial
discontinuity. 
We used the variable smoothing length
($\eta=1$, $C_{\rm smooth}=2.0$).

\subsection{Extreme Blast Wave}
                                                  \label{sec:BlastWave}

\begin{figure}[htp] 
\includegraphics[width=14.4cm]{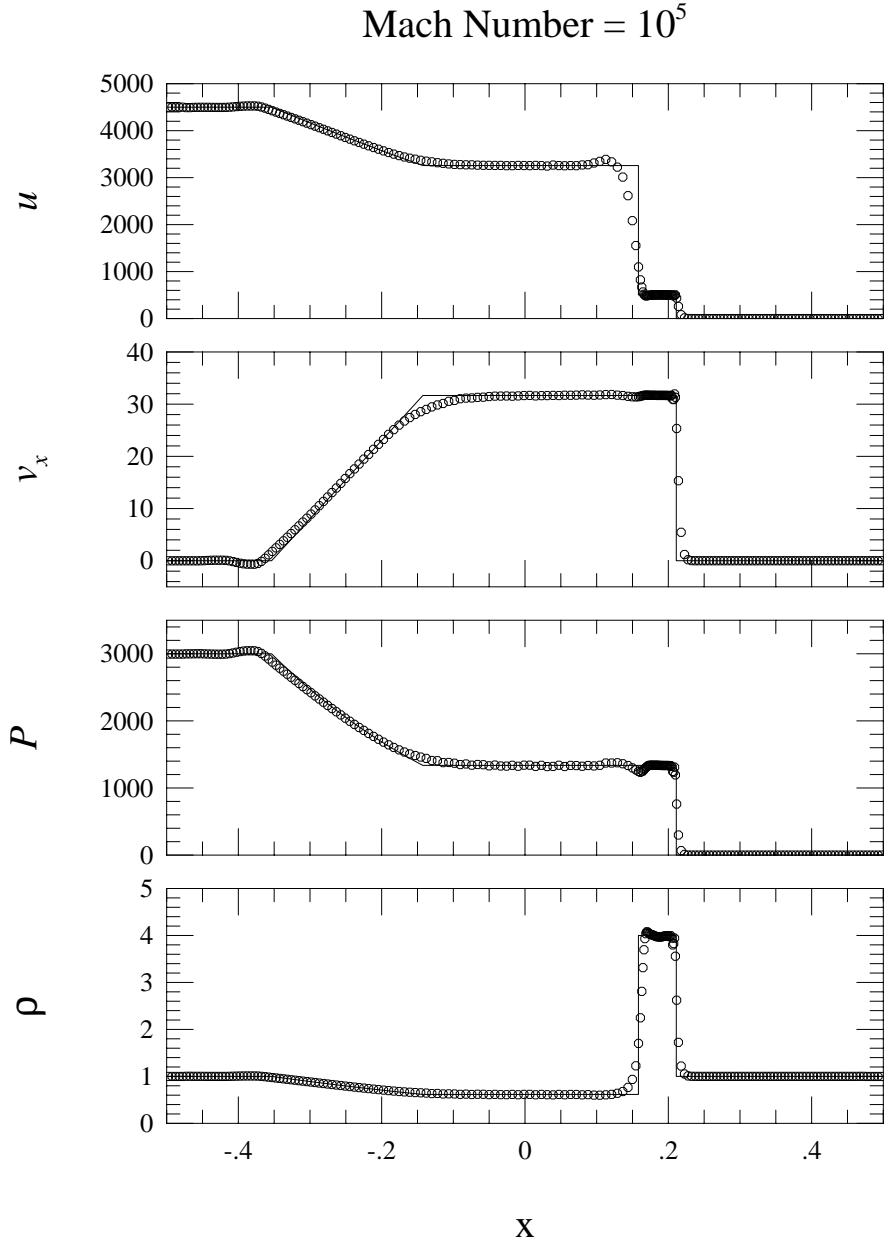}
   \caption
{
Results of 1D calculation for the extremely strong shock tube problem 
of which Mach number is about $10^5$. 
We used 100 particles on each side of the initial discontinuity.  
The open circles plot the result of the present method 
where the cubic spline approximation for the convolution 
Eq.(\ref{eq:V2ij}) and the variable smoothing length are used
($\eta=1$, $C_{\rm smooth}=2.0$). 
A small wiggle in the pressure distribution at the contact surface 
is due to the approximation for the convolution. 
                                                 \label{Fig:WC1} 
}
\end{figure}
\begin{figure}[htp] 
\includegraphics[width=14.4cm]{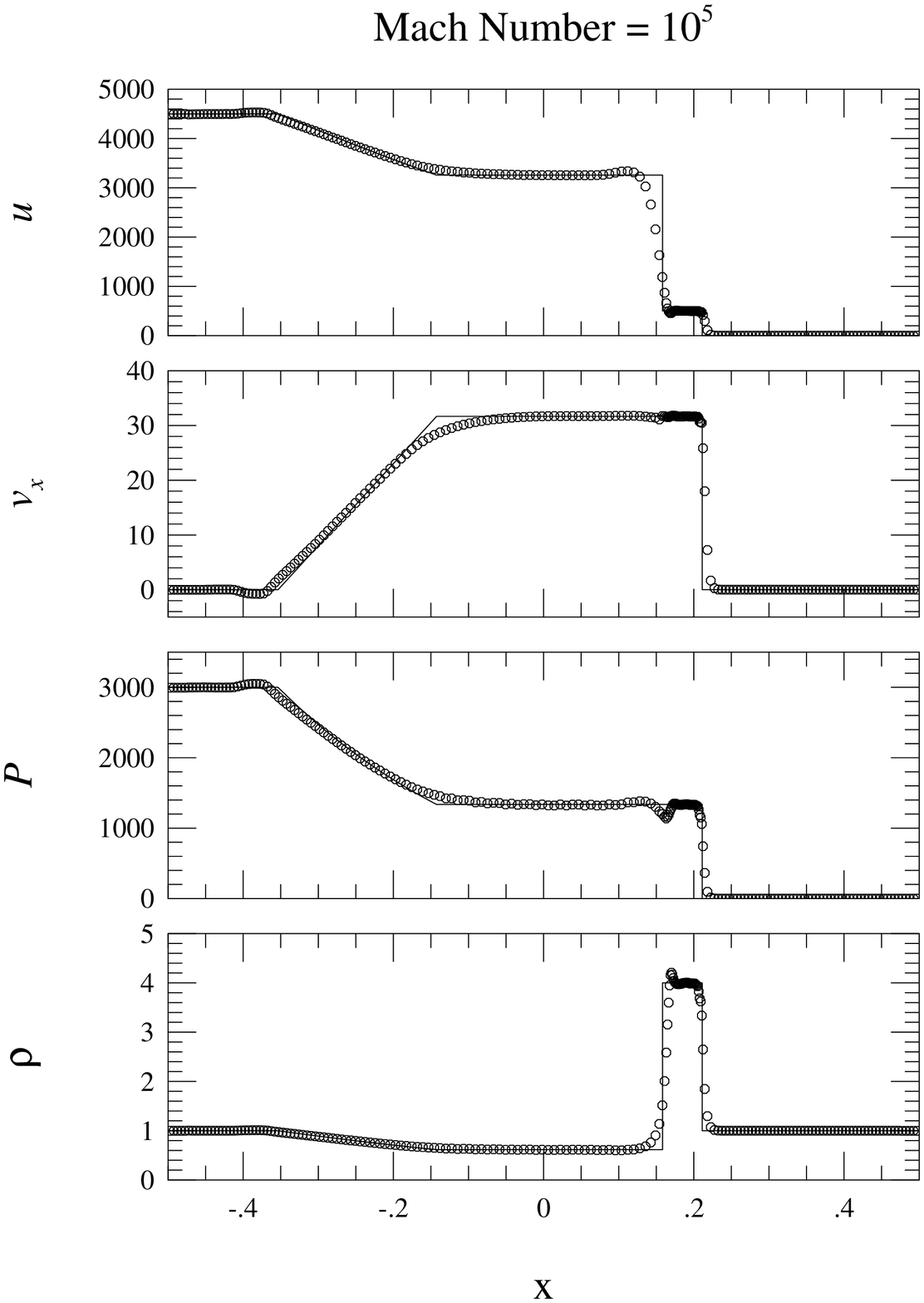}
   \caption
{
Same as Fig.7 except that we used the linear interpolation in 
the convolution Eq.(\ref{eq:V2ijL}).   
The amplitude of the pressure wiggling at the contact surface 
is slightly larger than that of the cubic spline result.  
                                                      \label{Fig:WCL} 
}
\end{figure}
To explore the capability of the present method, 
we test our method against an extremely strong shock tube problem. 
The initial parameters of the problem are the following:
\begin{eqnarray}
       \rho_{\rm L}  = 1 & , & \rho_{\rm R} = 1    , \nonumber \\
        P_{\rm L} = 3000 & , & P_{\rm R} = 10^{-7} , \nonumber \\
       v_{x,{\rm L}} = 0 & , & v_{x,{\rm R}} = 0   , 
\end{eqnarray}
where the initial pressure of the gas on the left-hand side is 
$3\times 10^{10}$ times that of the right-hand side. 
The ratio of specific heats is $\gamma = 5/3$.  
The Mach number is as large as $10^5$. 
We used 100 particles on each side of the initial discontinuity.  
Figure \ref{Fig:WC1} plots the result of the present method 
where the cubic spline approximation for the convolution 
Eq.(\ref{eq:V2ij}) and the variable smoothing length is used
($\eta=1$, $C_{\rm smooth}=2.0$). 
Even in this severe problem, 
the present method gave stable and accurate results and 
is free from penetration problems.
The pressure distribution shows a small wiggle at the 
contact surface.   
This is due to the approximation for the convolution. 
Figure \ref{Fig:WCL} plots the result of the present method with 
the linear interpolation Eq.(\ref{eq:V2ijL}).  
The amplitude of the pressure wiggling at the contact surface 
is slightly larger than that of the cubic spline result.  
In order to eliminate this unphysical wiggling, 
we need to develop more accurate approximation for the 
numerical convolution than Eq.(\ref{eq:V2ij}). 
Numerical experiments shows that the standard SPH cannot produce 
acceptable result at least with $\eta=1$, $\alpha \approx 1$, and 
$\beta \approx 2$.  
The possible fine-tuning of the parameters in the standard SPH 
may enable the calculation in this case.  
Note that the present method can describe this extreme phenomena 
with the same parameters ($\eta=1$, $C_{\rm smooth}=2.0$) 
used in the other test problems, 
without any tuning of the method. 

The accurate calculation with the single precision real numbers 
in the Fortran program is due to the conservative formulation  
with the internal specific energy $u$ instead of the total energy 
$e$ ($=v^2/2 + u$) or $E$ ($=\rho v^2/2 + \rho u$) which are 
usually used in the grid-based conservative numerical schemes  
(see Section \ref{Sec:Cons}).   
If we adopt $e$ or $E$ as the main variable for the integration, 
we have to do the subtraction to obtain $u$ ($=e-v^2/2$), 
which brings huge error into the numerical value of $u$ 
in the extremely supersonic motion 
(i.e., when $u \ll e \approx v^2/2$). 
In this sense our method has potential advantage 
over the grid-based higher-order Godunov methods, 
at least, in describing extremely supersonic flows.



\subsection{Wind-Tunnel}
To test the present method against multi-dimensional problem, 
we adopt the following wind-tunnel problem \cite{LPvL:15d},  
that can be in principle 
directly compared with the laboratory experiment. 
The geometry is a two-dimensional channel with a $15^{\circ}$ wedge 
on the lower wall. 
A $15^{\circ}$ expansion corner is also included.
The inflow Mach number is two. 
This kind of problem is poorly suited for the particle method 
because a rigid-wall boundary condition must be set up at 
the wall of the tunnel.  
We present this test problem only to demonstrate the 
capability of our scheme.

\begin{figure}[htp] 
\includegraphics[width=14.4cm]{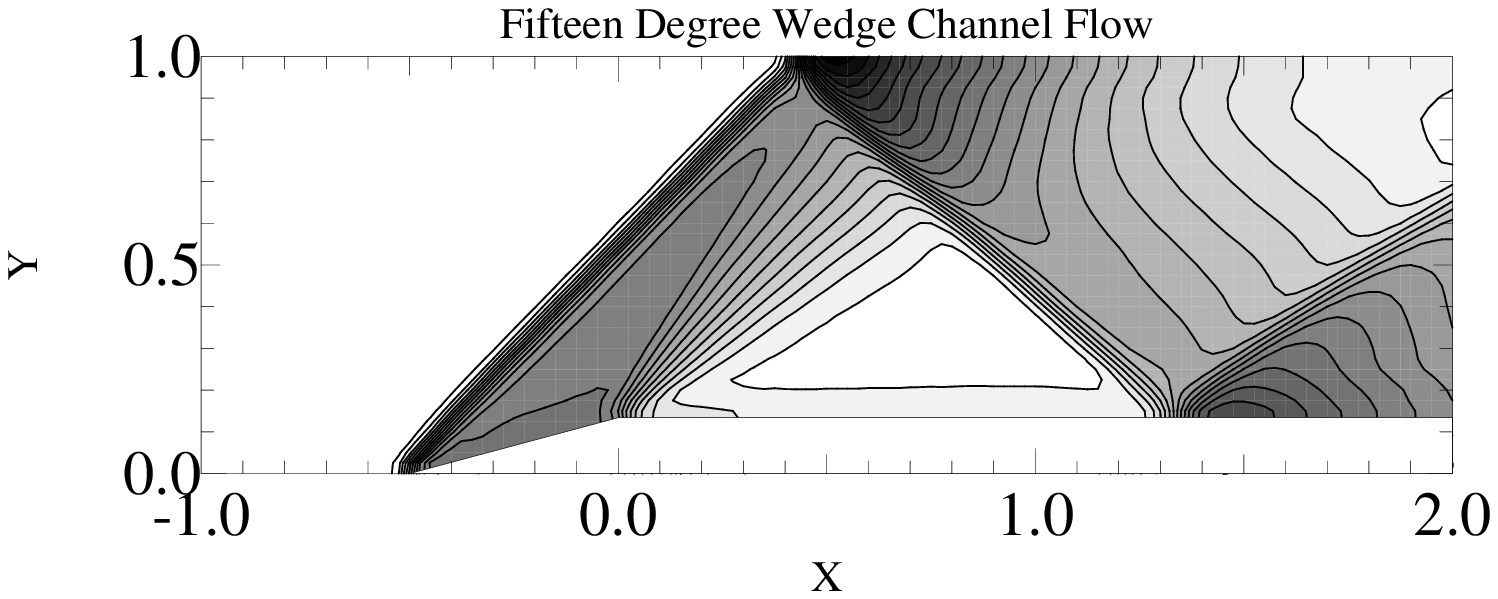}
   \caption
{
Results of calculation on ``fifteen degree wedge channel flow'' problem.
Initially 32$\times$96 particles are flowing inside the channel.
Mach-number contours from 0.92 to 1.97 with an increment of 0.05 
are plotted.
                                                    \label{Fig:15c}
}
\end{figure}

\begin{figure}[htp]
\includegraphics[width=14.4cm]{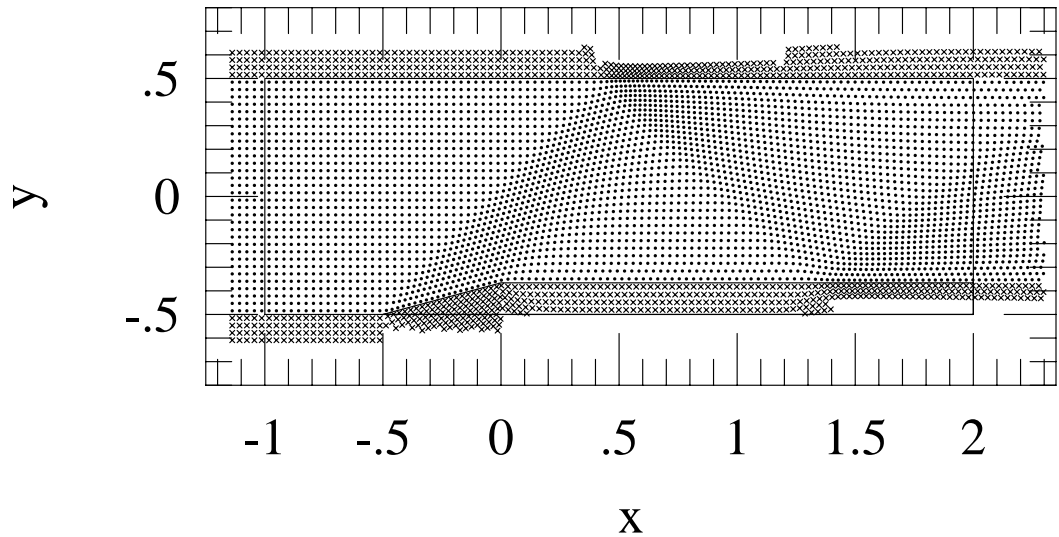}
   \caption
{
Same as Fig.\ref{Fig:15c} but positions of particles are plotted.
The crosses denote the positions of ``ghost particles'' that are 
introduced to mimic the rigid wall boundary condition.
                                                       \label{Fig:15p}
}
\end{figure}

\begin{figure}[htp] 
\includegraphics[width=14.4cm]{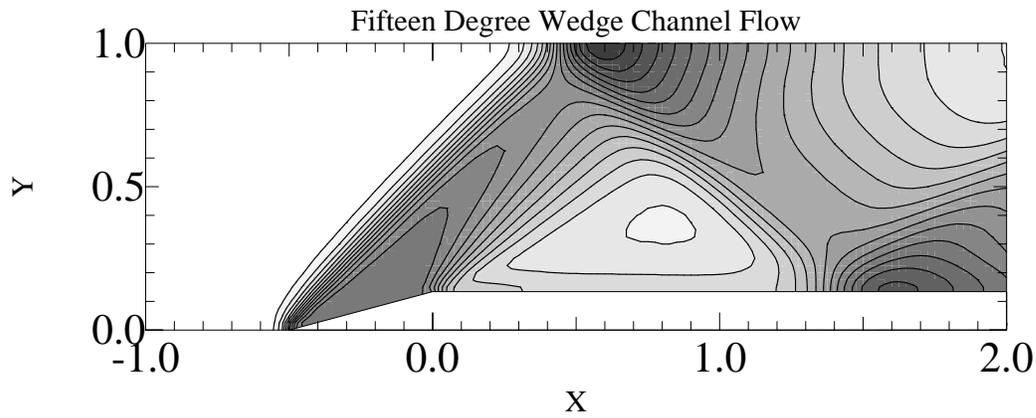}
   \caption
{
Results of calculation on ``fifteen degree wedge channel flow'' 
problem with the first-order Riemann Solver. 
Initially 32$\times$96 particles are flowing inside the channel.
The contour levels are the same as Fig.\ref{Fig:15c}.
                                                 \label{Fig:15cGSPH1st}
}
\end{figure}

\begin{figure}[htp] 
\includegraphics[width=14.4cm]{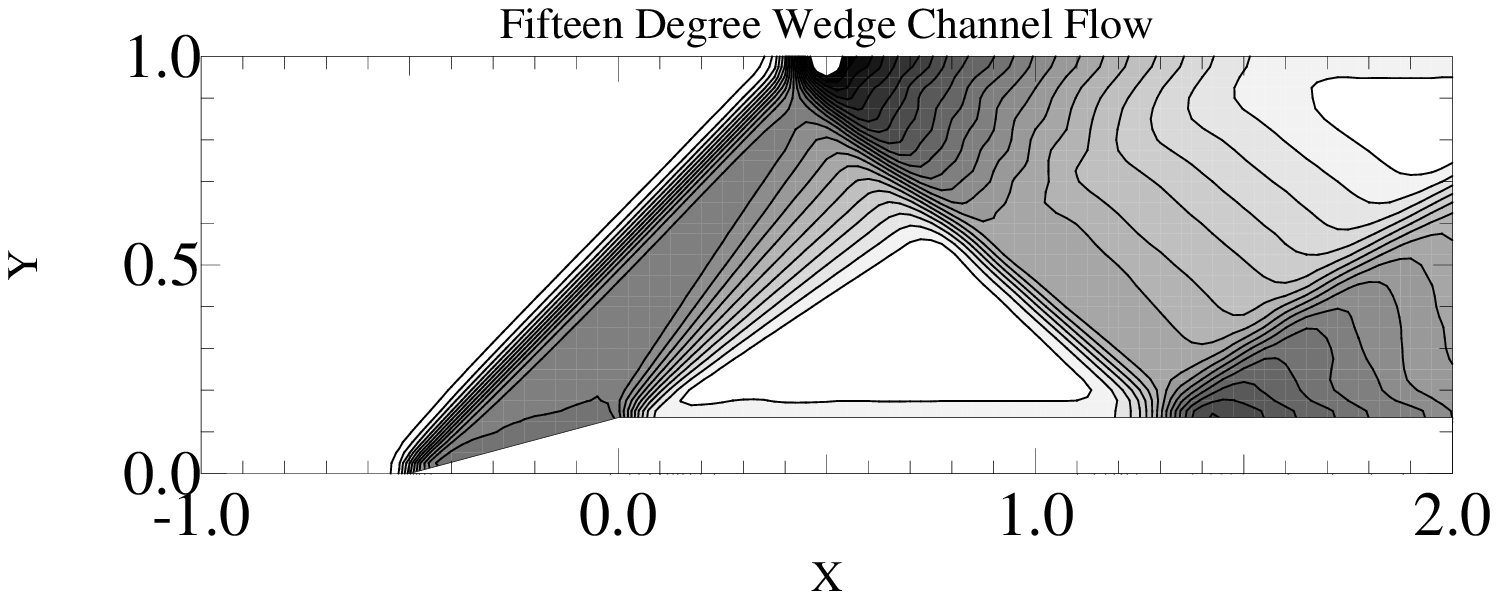}
   \caption
{
Results of calculation on ``fifteen degree wedge channel flow'' 
problem with the standard SPH.  
The standard artificial viscosity with $\alpha=1,\beta=2$ 
are used.  
Initially 32$\times$96 particles are flowing inside the channel.
The contour levels are the same as Fig.\ref{Fig:15c}.
                                                 \label{Fig:15cSSPHa1}
}
\end{figure}

\begin{figure}[htp] 
\includegraphics[width=14.4cm]{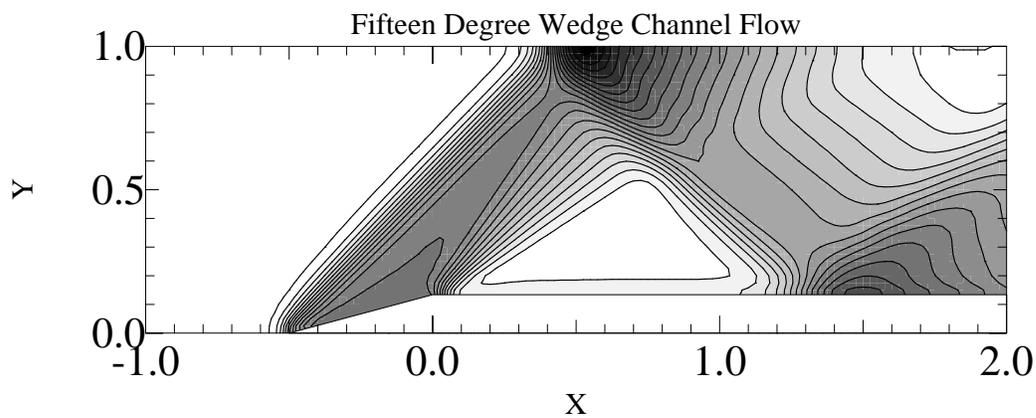}
   \caption
{
Results of calculation on ``fifteen degree wedge channel flow'' 
problem with the standard SPH.  
The standard artificial viscosity with $\alpha=2,\beta=4$ 
are used.  
Initially 32$\times$96 particles are flowing inside the channel.
The contour levels are the same as Fig.\ref{Fig:15c}.
                                                 \label{Fig:15cSSPHa2}
}
\end{figure}

Figure \ref{Fig:15c} shows our method's result. 
Mach-number contours from 0.92 to 1.97 with an increment of 0.05 
are plotted.  
Figure \ref{Fig:15p} shows particle positions. 
In this calculation, we realized the rigid-wall boundary condition 
by placing ``ghost particles'' in the wall. 
The inflowing particles 
outside of the left-hand side boundary 
were prepared at appropriate timesteps. 
The variable smoothing length is used
($\eta=1$, $C_{\rm smooth}=1.0$). 
Initially, 32$\times$96 particles were flowing inside the channel, 
which may be compared to the 32$\times$96 grids 
finite difference calculation in \cite{LPvL:15d}. 
Our results were satisfactory. 
For comparison, Figure \ref{Fig:15cGSPH1st} shows 
the result for the first-order Riemann Solver. 
The sharp features are smeared out.   

Figure \ref{Fig:15cSSPHa1} shows
results of calculation with the standard SPH, where   
the standard artificial viscosity with $\alpha=1,\beta=2$ 
are used.  
Note that the contour levels are the same as Fig. \ref{Fig:15c}. 
Owing to the moderate strength of the shocks in this problem, 
the standard SPH could produce reasonable result except for 
the region just after the Mach stem on the upper wall.  
A small hollow semicircle seen at $x \approx 0.5$, $y \approx 1$ 
corresponds to the small value of Mach number ($=0.806$) 
that is caused by an exceedingly large value of 
the post shock pressure.  
The calculations with larger values of 
the artificial viscosity parameters 
tend to remedy this pressure overshooting.  
Figure \ref{Fig:15cSSPHa2} shows the result of the standard SPH 
with $\alpha=2,\beta=4$ where the minimum value of Mach number 
behind the Mach stem is 0.957. 
In this case, however, the sharp features are smeared out. 
Even in this two dimensional problem of moderate Mach number, 
the present method shows its potential ability in analyzing 
supersonic flows.   

It is clear that a grid-based method must be used 
in the actual study of this kind of steady rigid-wall boundary problem. 
This result is presented only to demonstrate the 
capability of this particle scheme.
The present method becomes more useful when we study the hydrodynamical 
problems without rigid boundary that are common in 
astrophysics and space sciences.

\section{Summary}
                                                   \label{Sec:Summary}  
Smoothed Particle Hydrodynamics is reformulated by 
the formal convolution of the original hydrodynamics equations,  
and by a new action principle, in which the second order (in $h$) 
approximation is used for the kinetic term of the 
Lagrangian function.  
The force acting on each particle is determined by solving the 
Riemann problem for each particle pair. 
The prescription for variable smoothing length is also shown. 
These techniques are implemented in the strict conservation form. 
Numerical examples involving an extremely strong shock are shown. 
The other test calculations will be described elsewhere. 

Although the method with a spatially constant smoothing length is 
formulated in a rigorous manner, 
the method with the variable smoothing length is based on 
a crude approximation 
(Eqs. [\ref{eq:EoM6.5}],[\ref{eq:EoE6.5}]).  
A more refined technique for the variable smoothing length 
is to be studied. 
A better approximation than the cubic spline interpolation 
in the numerical convolution is required to eliminate 
the ``wiggle'' at the contact discontinuity in the blast wave 
problems (see Section \ref{sec:BlastWave}).

This paper presented a piece of concepts 
that are not discussed in detail in the literature. 
Those are 
the convolution of the original fluid equation 
(Eqs. [\ref{eq:EocEoM}],[\ref{eq:EocEoE}]), 
the definition and approximation for the velocity field 
(Eqs. [\ref{eq:Dov}], [\ref{eq:Aov}]), 
and the modification of the force due to dissipation 
(Eqs. [\ref{eq:EoM5}], [\ref{eq:EoE5}]). 
Incorporating these concept, 
the final evolution equation was cast into 
a form similar to the standard SPH equation. 
Therefore those concepts may enable the rigorous examination 
of the accuracy and stability of the method and 
may enable further modification.  

The numerical examples in Section \ref{sec:example} show 
that the present particle method based on Riemann Solver 
can handle severe problems with strong shocks, 
those might include the description of explosion/implosion
and supersonic jet phenomena.   
In this respect, further modification of the SPH method 
in modeling relativistic flows is promising with 
the help of relativistic Riemann Solver 
\cite{MartiMuller:1996}. 
In adittion, 
the Lagrangian particle methods have advantage over Eulerian 
grid-based methods, in describing chemically reacting 
(multi)fluid and radiatively heating/cooling fluid 
\cite{SI:NAP}. 
This is because we can simply assign 
the chemical composition and entropy to each particle 
as a fluid element.   
This direction has a wide area of applications.

\begin{acknowledgment}
The author thanks the anonymous referee for valuable comments. 
The author also thanks Toru Tsuribe, Yusuke Imaeda, and 
Shoken M. Miyama for useful discussions.  
\end{acknowledgment}

\appendix{}

\section{Derivation of the Equation of Motion}


In this Appendix we show the derivation of the equation of motion 
from the Euler-Lagrange's equation (Eq.[\ref{eq:EL}]). 
 \begin{eqnarray}
       \PD{L}{\Vec{x}_i} 
       & = &  - \sum_{k} m_k 
                         \int \PD{u}{\Vec{x}_i} 
                         W(\Vec{x}-\Vec{x}_k, h) d\Vec{x} \nonumber \\
       &   &  -          m_i 
                         \int u \PD{}{\Vec{x}_i} 
                         W(\Vec{x}-\Vec{x}_i, h) d\Vec{x} .
                                                     \label{eq:EL1}
 \end{eqnarray}
The first term on the RHS of this equation becomes the following:
 \begin{eqnarray}
       &   &  - \sum_{k} m_k 
                         \int \PD{u}{\Vec{x}_i} 
                         W(\Vec{x}-\Vec{x}_k, h) d\Vec{x} \nonumber \\
       & = &  - \sum_{k} m_k 
                         \int \frac{ P }{ \rho^2 }
                         \PD{ \rho }{ \Vec{x}_i }
                         W( \Vec{x}-\Vec{x}_k, h ) d\Vec{x} \nonumber \\
       & = &  - \sum_{k} m_k 
                         \int \frac{ P }{ \rho^2 }
                         m_i 
                         \PD{ W( \Vec{x}-\Vec{x}_i, h ) }{ \Vec{x}_i }
                         W( \Vec{x}-\Vec{x}_k, h ) d\Vec{x} .
 \end{eqnarray}

Before we manipulate the second term on the RHS of Eq. (\ref{eq:EL1}), 
we note that 
 \begin{equation}
              \PD{ W( \Vec{x}-\Vec{x}_i, h ) }{ \Vec{x}_i } 
         =  - \PD{ W( \Vec{x}-\Vec{x}_i, h ) }{ \Vec{x}   } .
 \end{equation}
We also obtain   
 \begin{equation}
              \int f(\Vec{x}) 
                   \PD{ W( \Vec{x}-\Vec{x}_i, h ) }{ \Vec{x} } 
                   d\Vec{x}
         =  - \int \PD{ f(\Vec{x}) }{ \Vec{x} } 
                   W( \Vec{x}-\Vec{x}_i, h ) 
                   d\Vec{x} ,
 \end{equation}
by the integration by parts and 
$W(\Vec{x}) \rightarrow 0 ~~(|\Vec{x}| \rightarrow \infty)$ . 

Using the above identities, we can transform the second term 
on the RHS of Eq. (\ref{eq:EL1}) as  
 \begin{eqnarray}
       &   &  -          m_i 
                         \int u \PD{}{\Vec{x}_i} 
                         W(\Vec{x}-\Vec{x}_i, h) d\Vec{x} \nonumber \\
       & = &  +          m_i 
                         \int u \PD{}{\Vec{x} } 
                         W(\Vec{x}-\Vec{x}_i, h) d\Vec{x} \nonumber \\
       & = &  -          m_i 
                         \int \PD{ u }{ \Vec{x} } 
                         W( \Vec{x}-\Vec{x}_i, h ) d\Vec{x} \nonumber \\
       & = &  -          m_i 
                         \int \frac{ P }{ \rho^2 }
                         \PD{ \rho }{ \Vec{x} }
                         W( \Vec{x}-\Vec{x}_i, h ) d\Vec{x} \nonumber \\
       & = &  -          m_i 
                         \int \frac{ P }{ \rho^2 }
                         \sum_{j} m_j
                         \PD{ W( \Vec{x}-\Vec{x}_j, h ) }{ \Vec{x} }
                         W( \Vec{x}-\Vec{x}_i, h ) d\Vec{x} \nonumber \\
       & = &             m_i 
                         \int \frac{ P }{ \rho^2 }
                         \sum_{j} m_j
                         \PD{ W( \Vec{x}-\Vec{x}_j, h ) }{ \Vec{x}_j }
                         W( \Vec{x}-\Vec{x}_i, h ) d\Vec{x}  .
 \end{eqnarray}

As a result, Euler-Lagrange's equation gives the following: 
 \begin{eqnarray}
                \ddot{\Vec{x}}_i
       & = &  - \sum_{j} m_j 
                         \int \frac{ P }{ \rho^2 }
                         \PD{}{\Vec{x}_i}
                         W(\Vec{x}-\Vec{x}_i, h)
                         W(\Vec{x}-\Vec{x}_j, h) d\Vec{x} \nonumber \\
       &   &  + \sum_{j} m_j 
                         \int \frac{ P }{ \rho^2 }
                         \PD{}{\Vec{x}_j}
                         W(\Vec{x}-\Vec{x}_i, h)
                         W(\Vec{x}-\Vec{x}_j, h) d\Vec{x} .
 \end{eqnarray}



\end{article}
\end{document}